\documentclass[aps, prb, nobibnotes, twocolumn, superscriptaddress, floatfix]{revtex4-2}
\usepackage{graphicx}
\usepackage{amsmath,amssymb}
\usepackage{booktabs}
\usepackage{xcolor}
\usepackage{amsfonts}
\usepackage{array}
\usepackage{mathrsfs}
\usepackage{bm}
\usepackage[title,titletoc]{appendix}
\usepackage{tabularx}
\usepackage{tikz}
\usetikzlibrary{arrows.meta, positioning, shapes.geometric, calc, backgrounds}
\usepackage{mathtools}

\newcolumntype{M}{>{$}c<{$}}
\renewcommand{\arraystretch}{2.2}

\usepackage[colorlinks=true, linkcolor=blue, citecolor=blue, urlcolor=blue]{hyperref}

\begin{document}

\title{Symmetry Criterion for Van Hove Criticality at Non-Time-Reversal-Invariant Momenta}

\author{Min-Quan Kuang}
\email{mqkuang@swu.edu.cn}
\affiliation{Chongqing Key Laboratory of Micro $\&$ Nano Structure Optoelectronics, School of Physical Science and Technology, Southwest University, Chongqing 400715, P. R. China}

\author{Hua-Yu Li}
\affiliation{Chongqing Key Laboratory of Micro $\&$ Nano Structure Optoelectronics, School of Physical Science and Technology, Southwest University, Chongqing 400715, P. R. China}

\begin{abstract}
At non-time-reversal-invariant momenta (non-TRIMs), time-reversal symmetry does not constrain the linear term of the band dispersion. Whether $\nabla E$ vanishes is therefore determined entirely by the representation theory of the little group. For nondegenerate bands, $\nabla E$ is forced to zero if and only if the vector representation $\Gamma_{\mathrm{vec}}$ of the little group does not contain the trivial representation $\Gamma_1$. When $\Gamma_{\mathrm{vec}}$ does contain $\Gamma_1$, $\nabla E$ is not forced to vanish for any nondegenerate band; the classification instead depends on the multiplicity of $\Gamma_1$ in $\Gamma_{\mathrm{vec}}$. For degenerate bands, the Wigner-Eckart theorem and Clebsch--Gordan coefficients determine whether linear couplings vanish, with classification performed at the subband level. Applied to space group 225, the criterion explains why the $W$ point is critical for all nondegenerate bands, the degenerate $E$ bands are generically noncritical, and the $K$ and $U$ points host parameter-dependent criticality. Supporting phase diagrams reveal a two-tier hierarchy: symmetry enforces $\nabla E=0$, while band parameters determine higher-order character. We extend this classification to all space groups hosting non-TRIMs in the single-group limit, providing a symmetry-dictated, parameter-independent framework for engineering Van Hove singularities in three-dimensional quantum materials.
\end{abstract}

\maketitle

\section{Introduction}

Van Hove singularities (VHSs) are nonanalytic structures in the electronic density-of-states (DOS) originating from critical points in the band structure where the group velocity vanishes. The resulting enhancement of the DOS near the Fermi level strongly amplifies electron-electron interactions, favoring collective phases such as superconductivity, magnetism, and charge density waves~\cite{vanhove1953, efremov2019, yuan2019, yuan2020, patra2025, classen2025}. In three dimensions, conventional VHSs are stationary points with $\nabla E=0$, classified by the Morse index into extrema ($M_0,M_3$) and saddle points ($M_1,M_2$), each exhibiting distinct DOS signatures~\cite{lifshitz1960, tamai2008, wu2021, tan2024}.

In our previous work~\cite{li2026}, we introduced a unified classification of three-dimensional VHSs that extends beyond conventional critical points to include noncritical types, where $\nabla E\neq 0$ along one direction. Critical VHSs are divided into ordinary ($M_0, M_1, M_2, M_3$) and higher-order ($T_1, T_2, T_3$) classes; noncritical VHSs fall into ordinary ($N_0, N_1, N_2$) and higher-order ($S_1, S_2$) classes. Through a tight-binding study of the pyrochlore lattice (space group 227), we found that the non-TRIMs $K=(3/8,3/8,3/4)$ and $U=(5/8,1/4,5/8)$ host the full noncritical spectrum---from $N_0,N_1,N_2$ to $S_1,S_2$---while $W=(1/2,1/4,3/4)$ hosts ordinary critical types $M_0,M_1,M_2,M_3$~\cite{li2026}. However, due to the multiple independent hopping parameters in the pyrochlore lattice, the phase diagrams in that work were presented as representative cross-sections of the full parameter space rather than complete two-dimensional phase diagrams. While these cross-sections demonstrated the existence of the predicted VHS types, they could not establish whether the critical versus noncritical dichotomy is truly universal across the entire parameter space or merely accidental in the chosen cross-sections.

The importance of non-TRIMs has been highlighted in recent studies. In SnTe, Dirac cones on (001)/(110) surfaces occur at non-TRIMs, in contrast to the (111) surface where they reside at TRIMs~\cite{liu2013}, indicating that surface orientation and little-group symmetry determine the location of Dirac points. Vortex Majorana modes further underscore this distinction~\cite{luo2025}. In Ta$_2$CS$_2$, non-TRIM $K$/$K'$ points exhibit valley-dependent Zeeman splitting under the little group $C_{3v}$~\cite{sarmah2025}. Unconventional Rashba splitting at non-TRIMs in Sn/SiC surfaces shows that Rashba-like spin textures can occur away from TRIMs~\cite{tao2023}. Most relevantly, phononic nodal chains require non-TRIM touching points; at TRIMs the chain reduces to a single nodal line~\cite{zhu2022}. These examples share a common theme: at non-TRIMs, time-reversal symmetry does not automatically enforce $\nabla E=0$, yet a systematic criterion for determining its vanishing has not been established. This motivates our symmetry-based criterion, which we apply to space group 225 and verify with complete phase diagrams.

Space group 225 ($Fm\bar{3}m$) serves as an ideal testing ground for two reasons. First, both space groups 225 and 227 are face-centered cubic structures, and the non-TRIMs $W$, $K$, and $U$ share identical little groups in both: $W$ has little group $D_{2d}$, while $K$ and $U$ have little group $C_{2v}$~\cite{bradley1972}. Second, an $s$-orbital model on the $4a$ Wyckoff position of space group 225 has only three independent hopping parameters, enabling complete two-dimensional phase diagrams that exhaustively cover the entire parameter space. Such complete phase diagrams are essential because they can definitively confirm whether the dichotomy between critical and noncritical behavior is enforced by symmetry rather than an artifact of fine-tuning. The complete phase diagrams obtained here thus serve as a direct and rigorous numerical verification of our symmetry predictions within this tight-binding framework.

Before presenting the criterion, we must distinguish two classes of high-symmetry points: time-reversal-invariant momenta (TRIMs) and non-TRIMs. A TRIM is defined by the condition $\mathbf{k}_0 \equiv -\mathbf{k}_0 \pmod{\mathbf{G}}$, i.e., there exists a reciprocal lattice vector $\mathbf{G}$ such that $\mathbf{k}_0 = -\mathbf{k}_0 + \mathbf{G}$; points for which no such $\mathbf{G}$ exists are non-TRIMs. At TRIMs, time-reversal (TR) symmetry enforces $E_n(\mathbf{k}_0+\mathbf{q}) = E_n(\mathbf{k}_0-\mathbf{q})$, which prohibits all odd-order terms in the expansion of the dispersion. Hence $\nabla E=0$ at TRIMs is a direct consequence of time-reversal invariance. The points $W$, $K$, and $U$ in the face-centered cubic Brillouin zone are precisely non-TRIMs. As a concrete example, $-W = (-1/2,-1/4,-3/4) \equiv (1/2,3/4,1/4)$ modulo reciprocal lattice vectors (taking the cubic lattice with lattice constant $a=1$), which is not equal to $W=(1/2,1/4,3/4)$; hence $W$ is a non-TRIM. The same verification applies to $K$ and $U$. At these points, time-reversal symmetry does not automatically enforce $\nabla E=0$, and the space-group symmetry alone determines the fate of the band gradient. Thus, non-TRIMs are precisely where a group-theoretical criterion---rather than a direct symmetry argument---is needed to determine whether $\nabla E$ vanishes, and they constitute the focus of this work.

%%%%%%%%
\begin{figure}[t]
\centering
\includegraphics[width=0.85\columnwidth]{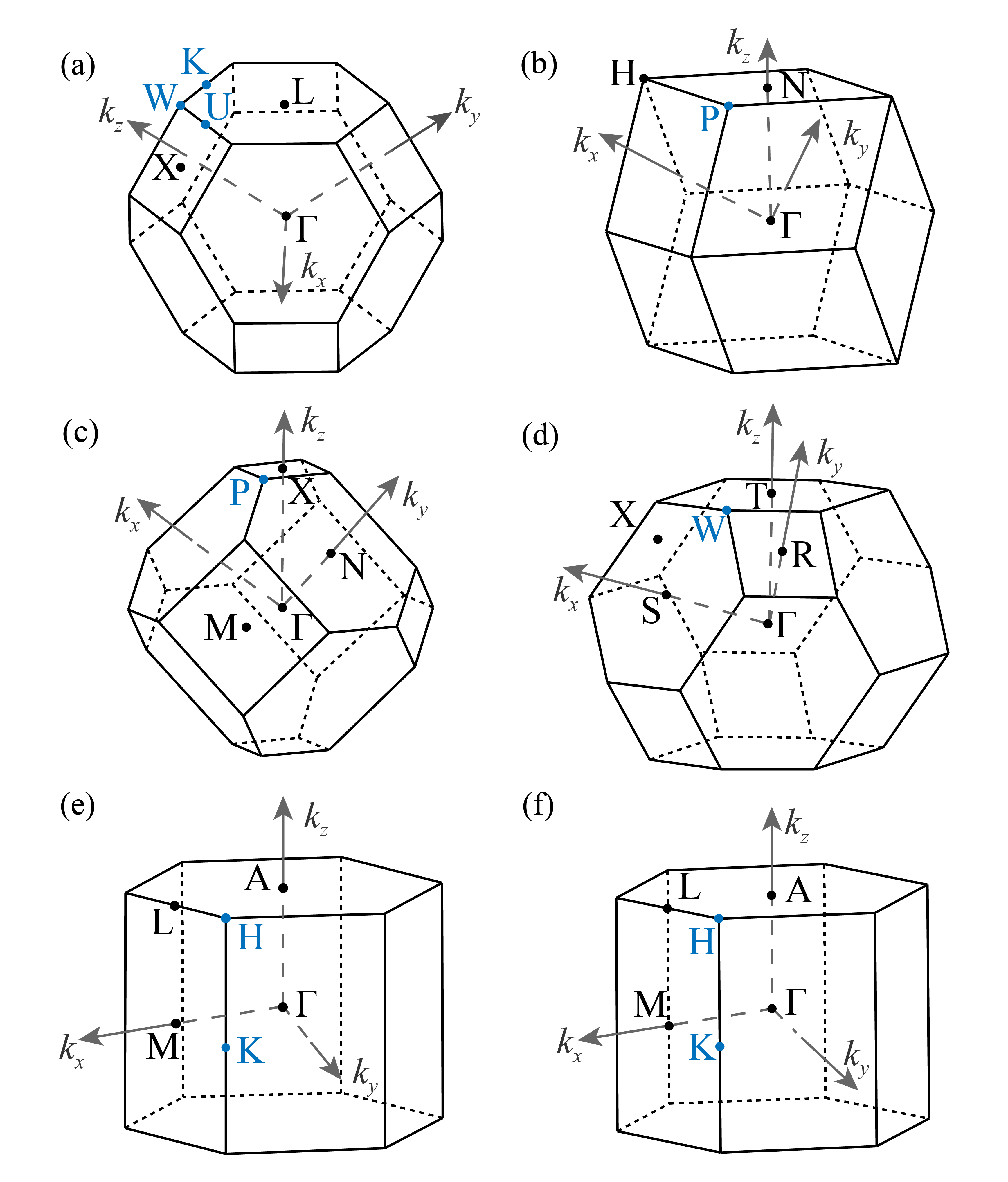}
\caption{Brillouin zones for the six crystal systems classified in Table~\ref{tab2}. The TRIMs and non-TRIMs are marked by black and blue dots, respectively. Panels: (a) face-centered cubic (FCC) lattice with TRIMs $\Gamma$, X, L and non-TRIMs W, K, U; (b) body-centered cubic (BCC) lattice with TRIMs $\Gamma$, H, N and non-TRIM P; (c) body-centered tetragonal (BCT) lattice with TRIMs $\Gamma$, M, X, N and non-TRIM P; (d) body-centered orthorhombic (BCO) lattice with TRIMs $\Gamma$, X, S, R, T and non-TRIM W; (e) hexagonal (Hex) lattice with TRIMs $\Gamma$, A, M, L and non-TRIMs K, H; and (f) trigonal (Trig) lattice in the hexagonal setting with TRIMs $\Gamma$, A, M, L and non-TRIMs K, H.}
\label{fig:bz}
\end{figure}
%%%%%%%%%%%%%

The use of group theory to classify critical points in band structures has a long history, dating back to the work of Herring~\cite{herring1937} and Bouckaert, Smoluchowski, and Wigner~\cite{bouckaert1936} on the symmetry of energy bands. More recently, the classification of topological semimetals~\cite{weng2015, bradlyn2016, armitage2018} and the characterization of nodal points~\cite{fang2016} have relied heavily on the representation theory of little groups. The question of whether a band gradient is forced to vanish by symmetry is closely related to the theory of $k\cdot p$ Hamiltonians and the method of invariants~\cite{luttinger1955, bir1974, winkler2003}. However, a systematic criterion for distinguishing critical from noncritical VHSs at non-TRIMs has not been previously established.

In this work, we establish a symmetry-based criterion that determines, purely from the little-group representation theory, whether $\nabla E$ is forced to vanish at non-TRIMs. For nondegenerate bands, the criterion is independent of the band's irreducible representation; for degenerate bands, it requires computing Clebsch--Gordan coefficients. Applied to space group 225, the criterion explains why the $W$ point is symmetry-enforced critical for all nondegenerate bands, while the $K$ and $U$ points are generically noncritical, with criticality appearing only upon parameter fine-tuning. These predictions are verified by complete tight-binding phase diagrams. We then extend the classification to all space groups containing non-TRIMs. The criterion applies to weakly correlated paramagnetic systems with full space-group symmetry. Generalizations to magnetically ordered or strong spin-orbit coupling (SOC) systems are discussed in Sec.~\ref{sec:II-E}.

%%%%%%%%%%%%%%%%%%%%%%%%%%%%%%
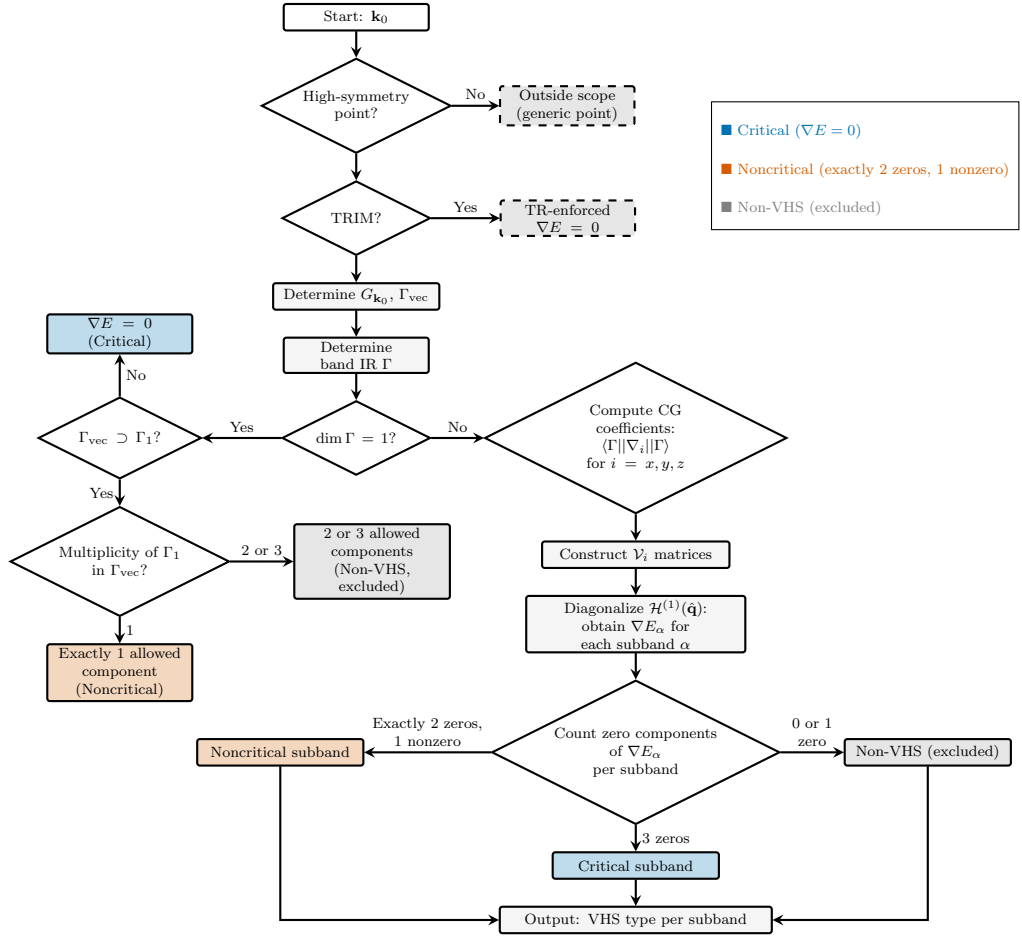
\begin{figure*}[t]
\centering
\begin{tikzpicture}[
    scale=0.7, transform shape,
    node distance=0.5cm,
    block/.style={rectangle, draw, thick, text width=2.6cm, align=center, 
                  minimum height=0.5cm, rounded corners=1pt, inner sep=2pt, font=\footnotesize},
    block-excl/.style={rectangle, draw, thick, text width=2.8cm, align=center, 
                  minimum height=0.5cm, rounded corners=1pt, inner sep=2pt, font=\footnotesize, fill=gray!20},
    decision/.style={diamond, draw, thick, text width=2.2cm, align=center, 
                     minimum height=0.5cm, inner sep=1pt, aspect=2, font=\footnotesize},
    arrow/.style={thick, ->, >=stealth},
]

% Color definitions
\definecolor{critColor}{RGB}{0, 114, 178}
\definecolor{noncritColor}{RGB}{213, 94, 0}

% ============================================================
% Panel 1: Initial checks (top)
% ============================================================
\node[block] (start) {Start: $\mathbf{k}_0$};

\node[decision, below=0.5cm of start] (highsym) {High-symmetry\\ point?};

\node[decision, below=0.5cm of highsym] (trim) {TRIM?};

% Right branches (outside scope)
\node[block, right=0.9cm of highsym, text width=2.4cm, fill=gray!20, font=\footnotesize, dashed]
    (generic) {Outside scope\\ (generic point)};

\node[block, right=1.3cm of trim, text width=2.4cm, fill=gray!20, font=\footnotesize, dashed]
    (trimtrue) {TR-enforced \\ $\nabla E=0$};

% Main path continues
\node[block, below=0.5cm of trim, text width=3.0cm, fill=gray!8] 
    (lgroup) {Determine $G_{\mathbf{k}_0}$, $\Gamma_{\text{vec}}$};

\node[block, below=0.5cm of lgroup, fill=gray!8] 
    (ir) {Determine band IR $\Gamma$};

\node[decision, below=0.5cm of ir] 
    (dim) {$\dim \Gamma = 1$?};

% ============================================================
% Panel 2: Nondegenerate bands (lower left)
% ============================================================
\node[decision, left=1.5cm of dim, text width=2.4cm] 
    (global) {$\Gamma_{\mathrm{vec}} \supset \Gamma_1$?};

% No branch -> all 1D bands are critical (upward)
\node[block, above=0.8cm of global, text width=2.6cm, fill=critColor!25, font=\footnotesize] 
    (global_no) {$\nabla E=0$\\ (Critical)};

% Yes branch -> check number of components containing Gamma (downward)
\node[decision, below=0.5cm of global, text width=2.8cm] 
    (count) {Multiplicity of $\Gamma_1$\\ in $\Gamma_{\mathrm{vec}}$?};

% Two branches (1 or 2/3)
\node[block, below=0.5cm of count, text width=2.6cm, fill=noncritColor!25, font=\footnotesize] 
    (count1) {Exactly 1 allowed\\ component\\ (Noncritical)};

\node[block-excl, right=1.2cm of count, text width=2.8cm, font=\footnotesize]
    (count2) {2 or 3 allowed\\ components\\ (Non-VHS, excluded)};

% ============================================================
% Panel 3: Degenerate bands (lower right)
% ============================================================
\node[decision, right=1.0cm of dim, text width=3.0cm] 
    (cg) {Compute CG coefficients:\\ $\langle \Gamma || \nabla_i || \Gamma \rangle$\\ for $i=x,y,z$};

\node[block, below=0.5cm of cg, text width=3.4cm, fill=gray!8, font=\footnotesize] 
    (construct) {Construct $\mathcal{V}_i$ matrices};

\node[block, below=0.5cm of construct, text width=4.0cm, fill=gray!8, font=\footnotesize] 
    (diag) {Diagonalize $\mathcal{H}^{(1)}(\hat{\mathbf{q}})$:\\ obtain $\nabla E_\alpha$ for each subband $\alpha$};

\node[decision, below=0.5cm of diag, text width=3.4cm] 
    (count_sub) {Count zero components\\ of $\nabla E_\alpha$\\ per subband};

% Three branches
\node[block, below=0.5cm of count_sub, text width=3.0cm, fill=critColor!25, font=\footnotesize] 
    (crit_sub) {Critical subband};

\node[block, left=2.4cm of count_sub, text width=3.0cm, fill=noncritColor!25, font=\footnotesize] 
    (noncrit_sub) {Noncritical subband};

\node[block-excl, right=1.2cm of count_sub, text width=3.0cm, font=\footnotesize]
    (excl_sub) {Non-VHS (excluded)};

% ============================================================
% Final output
% ============================================================
\node[block, below=0.5cm of crit_sub, text width=5.0cm, fill=gray!8, font=\footnotesize, draw=black, thick]
    (output) {Output: VHS type per subband};

% ============================================================
% Legend (placed in the upper-right blank area)
% ============================================================
\node[above right=6.0cm and 6.0cm of dim, font=\footnotesize, draw, fill=white, inner sep=3pt, anchor=north west] {
\begin{tabular}{l}
\color{critColor}$\blacksquare$ Critical ($\nabla E=0$) \\
\color{noncritColor}$\blacksquare$ Noncritical (exactly 2 zeros, 1 nonzero) \\
\color{gray}$\blacksquare$ Non-VHS (excluded)
\end{tabular}
};

% ============================================================
% Panel 1 internal arrows
% ============================================================
\draw[arrow] (start) -- (highsym);
\draw[arrow] (highsym) -- (trim);
\draw[arrow] (trim) -- (lgroup);
\draw[arrow] (lgroup) -- (ir);
\draw[arrow] (ir) -- (dim);

% ============================================================
% Panel 1 -> right branches (outside scope)
% ============================================================
\draw[arrow] (highsym.east) -- node[above, font=\footnotesize] {No} (generic.west);
\draw[arrow] (trim.east) -- node[above, font=\footnotesize] {Yes} (trimtrue.west);

% ============================================================
% Inter-panel arrows
% ============================================================
\draw[arrow] (dim.west) -- node[above, font=\footnotesize] {Yes} (global.east);
\draw[arrow] (dim.east) -- node[above, font=\footnotesize] {No} (cg.west);

% ============================================================
% Panel 2 internal arrows
% ============================================================
\draw[arrow] (global.south) -- node[left, font=\footnotesize] {Yes} (count.north);
\draw[arrow] (global.north) -- node[right, font=\footnotesize] {No} (global_no.south);

\draw[arrow] (count.south) -- node[right, font=\footnotesize] {1} (count1.north);
\draw[arrow] (count.east) -- node[above, font=\footnotesize] {2 or 3} (count2.west);

% ============================================================
% Panel 3 internal arrows
% ============================================================
\draw[arrow] (cg) -- (construct);
\draw[arrow] (construct) -- (diag);
\draw[arrow] (diag) -- (count_sub);

\draw[arrow] (count_sub.south) -- node[right, font=\footnotesize, align=center] {3 zeros} (crit_sub.north);
\draw[arrow] (count_sub.west) -- node[above, font=\footnotesize, align=center] {Exactly 2 zeros,\\ 1 nonzero} (noncrit_sub.east);
\draw[arrow] (count_sub.east) -- node[above, font=\footnotesize, align=center] {0 or 1 \\ zero} (excl_sub.west);

% ============================================================
% Panel 3 -> final output
% ============================================================
\draw[arrow] (crit_sub.south) -- (output.north);
\draw[arrow] (noncrit_sub.south) -- ++(0,-0.5) |- (output.west);
\draw[arrow] (excl_sub.south) -- ++(0,-0.5) |- (output.east);

\end{tikzpicture}
\caption{Flowchart for classifying VHS types at non-TRIM high-symmetry points. For nondegenerate bands ($\dim\Gamma=1$), the criterion proceeds in two steps: (i) check whether $\Gamma_{\mathrm{vec}}$ contains the trivial representation $\Gamma_1$; if not, all 1D bands are critical; (ii) if yes, count the multiplicity of $\Gamma_1$ in $\Gamma_{\mathrm{vec}}$: multiplicity 1 $\Rightarrow$ noncritical, multiplicity 2 or 3 $\Rightarrow$ Non-VHS (excluded). For degenerate bands ($\dim\Gamma>1$), each subband is classified independently by counting the zero components of $\nabla E_\alpha$: 3 zeros $\Rightarrow$ critical; exactly 2 zeros and 1 nonzero $\Rightarrow$ noncritical; otherwise (0 or 1 zero) $\Rightarrow$ Non-VHS (excluded). Dashed boxes indicate cases outside the scope of this work (generic points and TRIMs).}
\label{fig:flowchart}
\end{figure*}
%%%%%%%%%

\section{Symmetry criterion for the vanishing of the band gradient}
\label{sec:II}

\subsection{General formulation}
\label{sec:II-A}
In three-dimensional band structures, the DOS behavior of Van Hove singularities is determined by the local expansion of the band dispersion near the critical point. Let the energy dispersion of the $n$-th band near the high-symmetry point $\mathbf{k}_0$ be expanded as
\begin{equation}
E_n(\mathbf{k}_0+\mathbf{q}) = E_n(\mathbf{k}_0) + \sum_i v_i q_i + \frac{1}{2}\sum_{ij} H_{ij} q_i q_j + \mathcal{O}(q^3),
\label{eq:expansion}
\end{equation}
where $v_i = \partial E_n/\partial k_i|_{\mathbf{k}_0}$ are the components of the band gradient and $H_{ij} = \partial^2 E_n/\partial k_i \partial k_j|_{\mathbf{k}_0}$ are the Hessian matrix elements. 

If $\nabla E_n(\mathbf{k}_0) \equiv (v_x, v_y, v_z)$ vanishes, then $\mathbf{k}_0$ is a critical point and the DOS may exhibit nonanalytic behavior; if $\nabla E_n(\mathbf{k}_0) \neq 0$ with exactly one nonzero component, the point is a noncritical VHS~\cite{li2026}. Cases with two or three nonzero gradient components are excluded from the VHS classification. Determining whether the gradient vanishes is therefore the first step in the classification of VHSs.

The key question is whether this determination requires explicit band-structure calculations. As we show below, the answer is no: whether $\nabla E_n(\mathbf{k}_0)$ is forced to vanish is a symmetry property, determined solely by the little group of $\mathbf{k}_0$, independent of specific band parameters.

\subsection{Derivation of the criterion}
\label{sec:II-B}

Let $G_{\mathbf{k}_0}$ be the little group of $\mathbf{k}_0$, i.e., the subgroup of the space group that leaves $\mathbf{k}_0$ invariant modulo a reciprocal lattice vector~\cite{bouckaert1936, bradley1972}. For any symmetry operation $R \in G_{\mathbf{k}_0}$, the band energy satisfies
\begin{equation}
E_n(\mathbf{k}_0 + R\mathbf{q}) = E_n(\mathbf{k}_0 + \mathbf{q}).
\label{eq:symmetry}
\end{equation}
Substituting the expansion \eqref{eq:expansion} into \eqref{eq:symmetry} and comparing first-order terms yields
\begin{equation}
R\,\nabla E_n(\mathbf{k}_0) = \nabla E_n(\mathbf{k}_0), \qquad \forall R \in G_{\mathbf{k}_0}.
\label{eq:invariant}
\end{equation}
That is, the gradient vector must be invariant under all operations of the little group. The existence of a nonzero solution to Eq.~\eqref{eq:invariant} is governed by the representation content of $G_{\mathbf{k}_0}$~\cite{tinkham1964}: a nonzero invariant vector exists if and only if the trivial representation $\Gamma_1$ ((identity representation) ) appears in the vector representation $\Gamma_{\mathrm{vec}}$.

An ordinary polar vector in three-dimensional space transforms under a point-group operation $R$ as $\mathbf{v} \mapsto R\mathbf{v}$. This three-dimensional representation is the vector representation, denoted $\Gamma_{\mathrm{vec}}$. Decomposing $\Gamma_{\mathrm{vec}}$ into irreducible representations of the little group gives
\begin{equation}
\Gamma_{\mathrm{vec}} = \bigoplus_i \Gamma_i.
\label{eq:decomp}
\end{equation}
The criterion follows directly from this decomposition.

For a nondegenerate band belonging to the one-dimensional irreducible representation $\Gamma$, the linear coupling between the band and the gradient is governed by the matrix element $\langle u_n | \nabla H | u_n \rangle$, where $|u_n\rangle$ is the Bloch state. The existence of a nonzero matrix element is determined by the Wigner-Eckart theorem. Crucially, because the bra state $\langle u_n |$ transforms according to the complex conjugate representation $\Gamma^*$, the selection rule is governed by the tensor product:
\begin{equation}
\Gamma^* \otimes \Gamma_{\text{vec}} \otimes \Gamma \supset \Gamma_1.
\label{eq:wigner_general}
\end{equation}
For any one-dimensional representation (including complex conjugate pairs such as ${}^1E, {}^2E$ encountered in certain point groups), the identity $\Gamma^* \otimes \Gamma = \Gamma_1$ holds universally, because the squared modulus of any one-dimensional character is strictly unity ($|\chi_\Gamma(R)|^2 = 1$). Substituting this universal identity into Eq.~(\ref{eq:wigner_general}) yields:
\begin{equation}
\Gamma_{\text{vec}} \supset \Gamma_1.
\label{eq:we_reduced}
\end{equation}
This is the necessary and sufficient condition for a nondegenerate band to have a nonzero gradient: the vector representation of the little group must contain the trivial representation. Physically, since the band representation $\Gamma$ cancels out in the Wigner-Eckart reduction, this condition reduces to a purely geometric requirement: the point group at $\mathbf{k}_0$ must allow for the existence of a nonzero polar vector that is invariant under all symmetry operations, which is precisely the statement of Neumann's principle \cite{hahn2005} applied to the gradient.

It is crucial to recognize that Eq.~(\ref{eq:we_reduced}) is strictly independent of the specific irreducible representation $\Gamma$ of the band. This is a profound consequence of the intra-band nature of the expectation value $\langle u_n | \nabla H | u_n \rangle$: the band representation completely cancels out in the Wigner-Eckart reduction.

It is instructive to contrast this with the case of inter-band transitions (e.g., optical matrix elements $\langle \Gamma' | \nabla H | \Gamma \rangle$), where the initial and final states belong to different representations $\Gamma$ and $\Gamma'$. In that case, the Wigner-Eckart condition $\Gamma' \otimes \Gamma_{\mathrm{vec}} \otimes \Gamma \supset \Gamma_1$ does depend on the specific representations involved. This dependence on band representations is characteristic of inter-band processes. For the intra-band gradient considered here, however, the band representation cancels out entirely.

Thus, for nondegenerate bands, the criterion for a symmetry-enforced zero gradient is:
\begin{equation}
\nabla E_n(\mathbf{k}_0) \equiv 0 \;\;(\text{symmetry-enforced}) \quad \Longleftrightarrow \quad \Gamma_{\mathrm{vec}} \not\supset \Gamma_1,
\label{eq:criterion1_corrected}
\end{equation}
and the condition for a symmetry-allowed nonzero gradient is:
\begin{equation}
\nabla E_n(\mathbf{k}_0) \neq 0 \;\;(\text{symmetry-allowed}) \quad \Longleftrightarrow \quad \Gamma_{\mathrm{vec}} \supset \Gamma_1.
\label{eq:criterion2_corrected}
\end{equation}
Crucially, these criteria are independent of the band's irreducible representation $\Gamma$; they depend only on the little group $G_{\mathbf{k}_0}$ through its vector representation $\Gamma_{\mathrm{vec}}$.

For a degenerate band with a higher-dimensional irreducible representation $\Gamma$, the linear coupling is described by the matrix $\mathcal{H}^{(1)}$ introduced in Sec.~\ref{sec:II-D}. The condition for $\mathcal{H}^{(1)}$ to be \emph{possibly} nonzero is again given by the Wigner-Eckart theorem \eqref{eq:wigner_general}. However, as discussed in detail in Sec.~\ref{sec:II-D}, this is only a necessary condition; the actual vanishing of $\mathcal{H}^{(1)}$ requires computing the Clebsch--Gordan coefficients. The degenerate case thus requires additional analysis beyond the simple criterion \eqref{eq:criterion1_corrected}--\eqref{eq:criterion2_corrected}; this is the branch labeled ``compute CG coefficients'' in Fig.~\ref{fig:flowchart}. For higher-dimensional $\Gamma$, the condition $\Gamma \otimes \Gamma \neq \Gamma_1$ means that the band representation does not cancel out, and the CG coefficients must be computed to determine whether $\mathcal{H}^{(1)}$ vanishes for each gradient component.

When $\Gamma_{\mathrm{vec}}$ contains two or three $\Gamma_1$ components, multiple gradient components are symmetry-allowed, and the resulting bands fall outside the single-band VHS classification (Non-VHS, excluded), as summarized in Fig.~\ref{fig:flowchart} and discussed in Sec.~\ref{sec:IV}. 

The above derivation for nondegenerate bands follows directly from the Wigner-Eckart theorem~\cite{wigner1959,sakurai2020,tinkham1964}. For one-dimensional $\Gamma$, Eq.~\eqref{eq:wigner_general} reduces to the global condition $\Gamma_{\mathrm{vec}} \supset \Gamma_1$, as shown above. For higher-dimensional $\Gamma$, additional care is required: Eq.~\eqref{eq:wigner_general} determines only whether linear couplings are \emph{allowed} by symmetry; to determine whether they actually vanish, one must compute the Clebsch--Gordan coefficients. The degenerate case is discussed in detail in Sec.~\ref{sec:II-D}.

This criterion has a clear physical interpretation. At a high-symmetry point, the local environment imposes constraints on the possible forms of the band dispersion. The gradient $\nabla E_n$ describes the linear variation of energy with momentum. For such linear variation to exist, the gradient vector itself must be invariant under all symmetry operations of the little group, as expressed in Eq.~\eqref{eq:invariant}. This is possible if and only if the little group admits a nonzero invariant polar vector, i.e., $\Gamma_{\mathrm{vec}} \supset \Gamma_1$. Importantly, because the gradient is an intra-band property (the initial and final states are the same band), the band's own irreducible representation $\Gamma$ does not affect this condition; it cancels out in the Wigner-Eckart reduction. This is a key distinction from inter-band matrix elements, where the representations of both initial and final states enter the selection rules. If $\Gamma_{\mathrm{vec}}$ does not contain $\Gamma_1$, the linear term is symmetry-forbidden for all nondegenerate bands at $\mathbf{k}_0$, and the dispersion must be flat to first order. Importantly, this criterion is independent of material-specific parameters --- it depends only on the space-group symmetry and the position of $\mathbf{k}_0$ in the Brillouin zone. For nondegenerate bands, this completes the criterion. For degenerate bands, however, the irreducible representation $\Gamma$ must be identified to compute the Clebsch--Gordan coefficients, as we now discuss.

%%%%%%%%%%%%
\begin{table*}[t]
\caption{Criticality at non-TRIMs in space group 225 in the single-group limit. See Sec.~\ref{sec:III-A} for the group-theoretical analysis and Fig.~\ref{fig:flowchart} for the classification flow. Complete data for all space groups are given in Table~\ref{tab2}.}
\label{tab1}
\renewcommand{\arraystretch}{1.2}
\setlength{\tabcolsep}{4pt}
\centering
\begin{ruledtabular}
\begin{tabular}{c c c c}
Point & Band IR & $\nabla E=0$ forced? & VHS type(s) \\
\hline
$W$ & $A_1, A_2, B_1, B_2$ (1D) & Yes ($\Gamma_{\mathrm{vec}}\not\supset A_1$) & Critical (M/T) \\
$W$ & $E$ (2D) & No (coupling to $v_z$ allowed) & Param.-dep. \\
$K,U$ & $A_1, A_2, B_1, B_2$ (1D) & No ($v_z \sim A_1$ universally allowed) & Param.-dep. \\
\end{tabular}
\end{ruledtabular}
\footnotetext[1]{Param.-dep. = parameter-dependent; the actual type (Critical or Noncritical) is determined by specific band parameters, though they are generically noncritical (N/S-type). At the $W$ point, $\Gamma_{\mathrm{vec}} = B_2 \oplus E$ lacks $A_1$, strictly enforcing criticality for all 1D bands; however, the $E$ band symmetrized product contains $B_2$, permitting a nonzero $v_z$ gradient. At the $K$ and $U$ points, the $z$-component of the gradient transforms as $A_1$, providing a universal $v_z$ coupling for all 1D bands.}
\end{table*}

\subsection{Determination of the irreducible representation}
\label{sec:II-C}

Before applying the criterion to a specific band, one must determine its irreducible representation $\Gamma$ at $\mathbf{k}_0$. For nondegenerate bands this is not required for the critical/noncritical classification (Sec.~\ref{sec:II-B}), but for degenerate bands it is essential for computing the Clebsch--Gordan coefficients (Sec.~\ref{sec:II-D}). 

In practice, $\Gamma$ can be obtained from standard tabulated data such as the Bradley--Cracknell tables~\cite{bradley1972}, by projection operator techniques applied to Bloch eigenstates~\cite{tinkham1964}, or through automated tools including \textsc{SpaceGroupIrep}~\cite{spacegroupirep, bradlyn2017}, the Bilbao Crystallographic Server~\cite{bilbao, Aroyo2006a}, and \textsc{IRVSP}~\cite{GAO2021107760} or \textsc{PyProcar}~\cite{HERATH2020107080, LANG2024109063}. For magnetic or spin-orbit coupled systems, the \textsc{ToMSGKpoint} package provides the corresponding double-group and magnetic representations~\cite{tomsg}. In all cases, the assignment is determined by space-group symmetry and orbital character, not by hopping parameters.

\subsection{Application to degenerate bands}
\label{sec:II-D}

When the band at $\mathbf{k}_0$ is degenerate, $\Gamma$ is a higher-dimensional irreducible representation of the little group $G_{\mathbf{k}_0}$. In this case, the first-order correction to the band energy is no longer a simple scalar shift as in Eq.~\eqref{eq:expansion}; instead, it is described by an effective Hamiltonian matrix acting within the degenerate subspace:
\begin{equation}
\mathcal{H}^{(1)}(\mathbf{q}) = \sum_i \mathcal{V}_i q_i,
\label{eq:H1}
\end{equation}
where the $\mathcal{V}_i$ are Hermitian matrices acting on the degenerate states. The directional derivative of the energy along $\hat{\mathbf{q}}$ for each split subband $\alpha$ is given by the eigenvalues of $\mathcal{H}^{(1)}(\hat{\mathbf{q}})$:
\begin{equation}
\frac{\partial E_\alpha}{\partial q}(\hat{\mathbf{q}}) = \mathrm{eig}_\alpha\left[\mathcal{H}^{(1)}(\hat{\mathbf{q}})\right], \qquad \hat{\mathbf{q}} = \frac{\mathbf{q}}{|\mathbf{q}|},
\label{eq:eigen_grad}
\end{equation}
where the eigenvalues generically depend on the direction of approach $\hat{\mathbf{q}}$ in momentum space.

\subsubsection{Wigner-Eckart analysis: necessary vs. sufficient conditions}

The existence of nonzero matrix elements $\mathcal{V}_i$ is governed by the Wigner-Eckart theorem. As established in Sec.~\ref{sec:II-B}, the coupling between the degenerate states and the gradient operator requires the triple-product condition:
\begin{equation}
\Gamma \otimes \Gamma_{\mathrm{vec}} \otimes \Gamma \supset \Gamma_1.
\label{eq:wigner_degenerate}
\end{equation}

It is crucial to recognize that Eq.~\eqref{eq:wigner_degenerate} is a necessary condition, not a sufficient one. It states that the trivial representation appears in the triple product, which is required for any nonzero invariant coupling to exist. However, the actual matrix elements $\mathcal{V}_i$ are proportional to the Clebsch--Gordan (CG) coefficients:
\begin{equation}
\langle \Gamma_m | \nabla_i | \Gamma_n \rangle \propto \langle \Gamma_1 | \Gamma_m \otimes \Gamma_{\mathrm{vec}}^{(i)} \otimes \Gamma_n \rangle,
\label{eq:cg_matrix}
\end{equation}
where $\Gamma_{\mathrm{vec}}^{(i)}$ denotes the specific component of the vector representation. Even when Eq.~\eqref{eq:wigner_degenerate} is satisfied, the CG coefficients for a particular component $i$ or for a specific set of basis functions may vanish identically due to the detailed structure of the point group.

For the point groups considered in this work ($D_{2d}$, $C_{2v}$, and their subgroups), we have explicitly verified the CG coefficients using standard character tables and basis-function transformations. In these cases, the following scenarios arise:
\begin{itemize}
\item For one-dimensional representations, Eq.~\eqref{eq:wigner_degenerate} reduces to the condition that the vector representation must contain the trivial representation, as established in Sec.~\ref{sec:II-B}. This condition is independent of the specific one-dimensional representation $\Gamma$.
\item For two-dimensional representations (e.g., the $E$ representation of $D_{2d}$), the CG coefficients $\langle E || \nabla_i || E \rangle$ must be computed explicitly. At the $W$ point of space group 225, the symmetrized product expands as
\begin{equation}
[E \otimes E]_{\mathrm{sym}} = A_1 \oplus B_1 \oplus B_2 .
\label{eq:Esym_W}
\end{equation}
Since $\nabla_z \sim B_2$ is contained in this product, the $z$-component coupling is symmetry-allowed ($\mathcal{V}_z \neq 0$ generically); the $x$ and $y$ components transform as $E$ and are strictly forbidden ($\mathcal{V}_x = \mathcal{V}_y = 0$). Consequently, the $E$ bands are generically noncritical, with $\nabla E = (0, 0, \pm v)$ for the two subbands, where $v$ is determined by the reduced matrix element. Accidental criticality ($v = 0$) may occur for specific parameter values where the reduced matrix element vanishes.
\end{itemize}

\subsubsection{Subband-resolved classification}

For a degenerate band, the classification must be performed at the level of individual subbands rather than for the degenerate multiplet as a whole. This is because different subbands within the same irreducible representation can have distinct gradient vectors, as determined by the CG coefficients. The procedure is as follows:

\begin{enumerate}
\item \textbf{Determine the little group} $G_{\mathbf{k}_0}$ and its irreducible representations.
\item \textbf{Compute the CG coefficients} $\langle \Gamma || \nabla_i || \Gamma \rangle$ for all three gradient components $i = x, y, z$, using the explicit basis functions of the degenerate subspace.
\item \textbf{Construct the $\mathcal{V}_i$ matrices} from the CG coefficients.
\item \textbf{Diagonalize $\mathcal{H}^{(1)}(\hat{\mathbf{q}}) = \sum_i \mathcal{V}_i \hat{q}_i$} to obtain the gradient vector $\nabla E_\alpha = (v_{\alpha,x}, v_{\alpha,y}, v_{\alpha,z})$ for each subband $\alpha$.
\item \textbf{Classify each subband independently} by counting the number of zero components of $\nabla E_\alpha$:
  \begin{itemize}
  \item If all three components are zero ($v_{\alpha,x} = v_{\alpha,y} = v_{\alpha,z} = 0$), the subband is \textbf{critical}.
  \item If exactly two components are zero and one is nonzero, the subband is \textbf{noncritical}.
  \item Otherwise (zero or one zero component), the subband does not meet the criteria for either critical or noncritical VHSs and is labeled as \textbf{Non-VHS (excluded)}. These cases are not classified as VHSs in this work.
  \end{itemize}
\item \textbf{Output} a list of types for all subbands in the degenerate multiplet.
\end{enumerate}

This subband-resolved classification is essential because different subbands within the same degenerate multiplet can have distinct gradient vectors. For example, in a two-dimensional representation, the two subbands may have gradient vectors $(v, 0, 0)$ and $(0, v, 0)$, respectively, leading to different classifications. The CG coefficients encode this information through the specific matrix structure of $\mathcal{V}_i$. The procedure is summarized in the right branch of the flowchart in Fig.~\ref{fig:flowchart}.

\subsubsection{Direction dependence and the definition of criticality}

A subtle but important point is that even if $\mathcal{H}^{(1)}(\hat{\mathbf{q}})$ has zero eigenvalues for some directions $\hat{\mathbf{q}}$, this does \emph{not} imply that the subband is critical. The DOS divergence at a critical point requires the gradient to vanish identically for all directions in the immediate neighborhood of $\mathbf{k}_0$. If $\partial E_\alpha/\partial q(\hat{\mathbf{q}}) = 0$ only along a specific direction (or a set of directions) while remaining nonzero for generic directions, the low-energy dispersion is still linear along most directions, and the DOS does not exhibit a singularity.

Mathematically, for a given subband $\alpha$, this condition is expressed as:
\begin{equation}
\text{Subband } \alpha \text{ is critical} \iff v_{\alpha,x} = v_{\alpha,y} = v_{\alpha,z} = 0.
\label{eq:critical_iff_subband}
\end{equation}
If any component $v_{\alpha,i} \neq 0$, the subband is noncritical because the linear term dominates the low-energy dispersion along at least one direction. If the subband has zero or one zero component, it is classified as Non-VHS and excluded from the present classification. This distinction is captured in the flowchart by the decision diamond ``For each subband $\alpha$: count zero components of $\nabla E_\alpha$.''

\subsection{Role of spin-orbit coupling and symmetry-lowering transitions}
\label{sec:II-E}

The analysis in Secs.~\ref{sec:II-B}--\ref{sec:II-D} assumes the single-group limit, where SOC is negligible and the electron spin is a good quantum number. In materials with strong SOC, however, the relevant symmetry group must be extended to the double group. The double group accounts for the transformation of spinor wavefunctions under $2\pi$ rotations: a $2\pi$ rotation, which is the identity in the single group, acts as $-1$ on a spin-$1/2$ spinor and is therefore represented by a  distinct (or non-identity) group element in the double group~\cite{winkler2003, dresselhaus2008}.

This extension has two important consequences for our criterion:
\begin{enumerate}
\item \textbf{The irreducible representation changes:} $\Gamma$ must be replaced by its double-group counterpart $\Gamma^D$. A band that is nondegenerate in the single group may become part of a Kramers-degenerate doublet in the double group, or vice versa. For example, in space group 225 at the $\Gamma$ point, the single-group $T_{2g}$ representation becomes the double-group $\Gamma_7$ or $\Gamma_8$ representation, depending on the orbital character and the strength of SOC~\cite{bradley1972}. This change in $\Gamma$ can alter whether the CG coefficients vanish.
\item \textbf{The vector representation decomposition changes:} In the double group, the vector representation $\Gamma_{\mathrm{vec}}$ must be compatible with the transformation of spinor wavefunctions. Additional representations may appear in its decomposition, and the CG coefficients must be recomputed.
\end{enumerate}

Consequently, a point that is critical in the single-group limit may become noncritical in the presence of strong SOC, and \emph{vice versa}. The classification presented in Table~\ref{tab2} applies strictly to the single-group limit; for materials with strong SOC, the double-group representations must be used instead, and the classification may differ.

Similarly, if the system undergoes a symmetry-lowering phase transition --- such as magnetic ordering that reduces the space group to a magnetic subgroup --- the irreducible representation $\Gamma$ and the little group $G_{\mathbf{k}_0}$ must be updated accordingly. In such cases, the CG coefficients must be recomputed using the appropriate magnetic little group or double-group little group. 

The mathematical structure of the criterion --- counting the number of zero components of $\nabla E_\alpha$ for each subband --- remains universal regardless of whether one works in the single group, double group, or magnetic group. Its predictive power, however, depends on correctly identifying the symmetry group of the \emph{actual} ground state. This flexibility makes the framework applicable to a wide range of material systems, from weakly correlated paramagnets to strongly spin-orbit-coupled or magnetically ordered compounds.

\section{Case study: Space group 225}
\label{sec:III}

\subsection{Group-theoretical predictions}
\label{sec:III-A}

We now apply the criterion established in Sec.~\ref{sec:II} to space group 225 ($Fm\bar{3}m$). This space group contains three non-TRIM high-symmetry points---$W$, $K$, and $U$---with distinct little groups, providing an ideal testing ground for the criterion. Character tables and representation labels follow the conventions of Bradley and Cracknell~\cite{bradley1972}. Figure~\ref{fig:bz} shows the face-centered cubic Brillouin zone and the locations of these points. All results in this section are for the single-group limit; generalization to strong SOC is discussed in Sec.~\ref{sec:II-E}.

\paragraph{$W$ point:} coordinates $(1/2, 1/4, 3/4)$. The little group is $D_{2d}$ ($\bar{4}2m$). The vector representation decomposes as
\begin{equation}
\Gamma_{\text{vec}} = B_2 \oplus E.
\label{eq:Wvec}
\end{equation}
Since $\Gamma_{\text{vec}}$ does not contain the trivial representation $A_1$, the criterion of Sec.~\ref{sec:II-B} implies that \emph{all} nondegenerate bands at $W$ have $\nabla E=0$ forced by symmetry, independent of their irreducible representation. Specifically, for the one-dimensional representations $A_1$, $A_2$, $B_1$, and $B_2$,
\begin{equation}
\Gamma_{\text{vec}} \not\supset A_1 \;\Longrightarrow\; v_x = v_y = v_z = 0,
\end{equation}
so all nondegenerate bands at $W$ are critical.

For the two-dimensional $E$ representation, the selection rule is determined by the symmetrized product $[E \otimes E]_{\text{sym}}$, as the first-order Hamiltonian must be Hermitian. For $D_{2d}$,
\begin{equation}
[E \otimes E]_{\text{sym}} = A_1 \oplus B_1 \oplus B_2.
\end{equation}
The $z$-component of the gradient transforms as $B_2$, which is contained in $[E \otimes E]_{\text{sym}}$; by the Wigner-Eckart theorem, $\langle E_m | \nabla_z | E_n \rangle$ is symmetry-allowed. The $x$ and $y$ components transform as $E$, absent from the symmetrized product, so $v_x = v_y = 0$ strictly. Thus the $E$ band has exactly two zero gradient components and one generically nonzero component ($v_z \neq 0$), making it noncritical in the generic parameter space.

\paragraph{$K$ and $U$ points:} These have coordinates $K = (3/8, 3/8, 3/4)$ and $U = (5/8, 1/4, 5/8)$, with little group $C_{2v}$ ($mm2$). The vector representation decomposes as
\begin{equation}
\Gamma_{\text{vec}} = A_1 \oplus B_1 \oplus B_2.
\label{eq:Kvec}
\end{equation}
Here $A_1$ corresponds to the $z$-component of the gradient ($\nabla_z \sim A_1$), while $x$ and $y$ components transform as $B_1$ and $B_2$, respectively.

For any one-dimensional representation $\Gamma$, the intra-band gradient matrix element $\langle u_n | \nabla_i | u_n \rangle$ requires
\begin{equation}
\Gamma \otimes \Gamma_{\mathrm{vec}}^{(i)} \otimes \Gamma = \Gamma_{\mathrm{vec}}^{(i)} \supset A_1 \;\Longleftrightarrow\; \Gamma_{\mathrm{vec}}^{(i)} = A_1.
\end{equation}
Thus the existence of a nonzero gradient component is independent of $\Gamma$ and depends only on whether that component transforms as $A_1$. Applying this to $C_{2v}$: the $z$-component ($A_1$) is symmetry-allowed, while the $x$ and $y$ components ($B_1, B_2$) are strictly forbidden. Consequently, for all one-dimensional (1D) bands ($A_1, A_2, B_1, B_2$), the gradient is constrained to $(0, 0, v_z)$, with exactly two zero components and one generically nonzero component. Hence all one-dimensional bands at $K$ and $U$ are noncritical (N/S-type).

We emphasize that even when the Hessian contains saddle-like cross terms such as $\alpha q_x q_y$, yielding eigenvalues $\alpha, -\alpha, 0$, the nonzero linear term $\nabla E$ dominates the low-energy behavior and precludes a DOS divergence~\cite{li2026}. The $K$ and $U$ points thus remain universally noncritical across all one-dimensional representations.

\paragraph{Symmetry-enforced vs. accidental criticality:}
A gradient that is strictly forbidden by symmetry must be distinguished from one that vanishes only for specific parameter combinations. At $W$, any 1D band is symmetry-enforced critical: $\nabla E=0$ throughout the entire parameter space, which is why the corresponding phase diagram [Fig.~3(a)] contains exclusively critical phases. In contrast, for the $E$ band at $W$ and all 1D bands at $K$ and $U$, symmetry permits a nonzero $v_z$, rendering them generically noncritical. However, parameter tuning can accidentally drive $v_z$ to zero; this manifests as lower-dimensional boundaries---critical lines in the predominantly noncritical phase diagram for the $K$ point [Fig.~3(b)].

Table~\ref{tab1} summarizes the criticality classification for all irreducible representations at these three non-TRIMs. The complete data for all space groups are given in Table~\ref{tab2}.

\subsection{Tight-binding phase diagram verification}
\label{sec:III-B}

\begin{figure}[ht]
\centering
\includegraphics[width=\columnwidth]{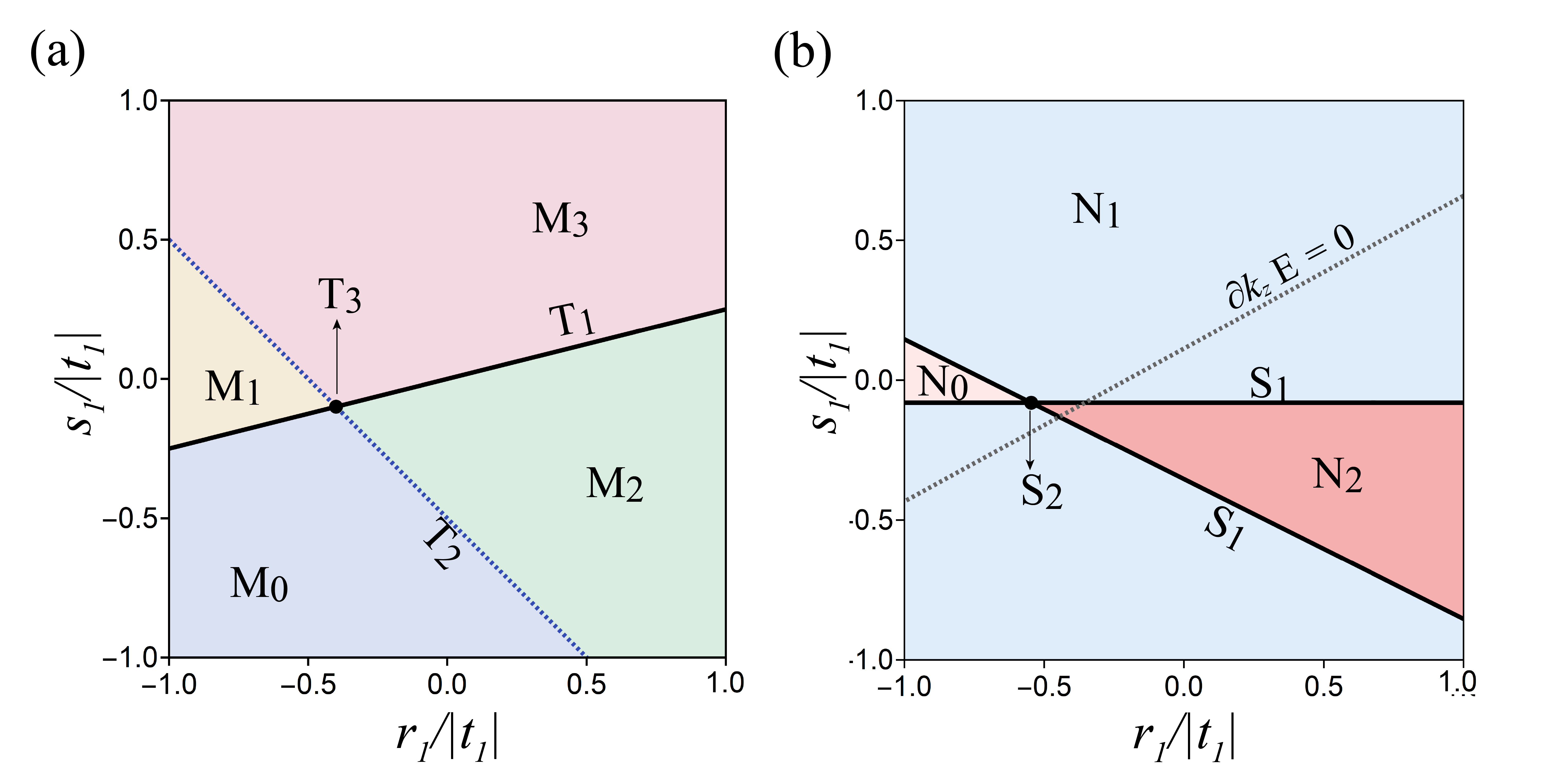}
\caption{Tight-binding phase diagrams for the $s$-orbital model in the single-group limit (excluding spin-orbit coupling). (a) Phase diagram at the $W$ point ($A_1$ representation), showing exclusively critical phases ($M_0$--$M_3$, $T_1$--$T_3$) as enforced by symmetry ($\Gamma_{\mathrm{vec}}\not\supset A_1$). The higher-order phase $T_1$ corresponds to the solid black line, while $T_2$ is denoted by the dotted blue line. (b) Phase diagram at the $K$ point, dominated by noncritical phases ($N_0$--$N_2$, $S_1, S_2$). The higher-order noncritical phase $S_1$ is explicitly marked by the solid black lines. True criticality emerges only accidentally along the gray dotted line ($\partial_{k_z} E=0$) via specific parameter tuning. In both panels, the shaded background colors delineate the ordinary phases ($M_0$--$M_3$ and $N_0$--$N_2$) which occupy finite parameter regions. The highest-order singularities, $T_3$ and $S_2$, emerge as black dots at the exact intersections of these respective boundary lines.}
\label{fig:phasediagram}
\end{figure}

To verify the group-theoretical predictions, we construct a minimal tight-binding model on the $4a$ Wyckoff position of space group~225 using spinless $s$-orbitals in the single-group (no SOC) limit. The model includes nearest-neighbor ($t_1$), second-neighbor ($r_1$), and third-neighbor ($s_1$) hoppings. The analytical band dispersion and the algorithmic VHS classification procedure are provided in Appendix~\ref{sec:TB225}. By exhausting the parameter space $(r_1/|t_1|, s_1/|t_1|)$ with $t_1=-1$ as the energy unit, we obtain complete two-dimensional phase diagrams for the $W$ and $K$ points (Fig.~\ref{fig:phasediagram}). Unlike representative cross-sections\cite{li2026}, these complete phase diagrams exhaustively cover all possible hopping-parameter combinations, ruling out any accidental fine-tuning as the origin of the observed behavior.

We begin with the $W$ point, where $\Gamma_{\mathrm{vec}}\not\supset A_1$. As shown in Fig.~\ref{fig:phasediagram}(a), the entire $(r_1,s_1)$ plane consists exclusively of critical VHS phases ($M_0$--$M_3$ and $T_1$--$T_3$, see legend). The hopping parameters control only the Hessian eigenvalues, driving Lifshitz transitions among ordinary saddle points ($M_1,M_2$, occupying finite regions), extrema ($M_0,M_3$), and higher-order critical lines or points ($T_1$--$T_3$) where $\det\mathcal{H}=0$. Crucially, $\nabla E$ remains strictly zero across the entire diagram, exactly as required by symmetry.

The $K$ point presents a starkly different picture. Here $\Gamma_{\mathrm{vec}}\supset A_1$, and the gradient is therefore symmetry-allowed rather than forbidden. Accordingly, Fig.~\ref{fig:phasediagram}(b) shows that the vast majority of the parameter space is occupied by noncritical VHS phases ($N_0$--$N_2$ and $S_1,S_2$). Critical phases ($M$- and $T$-types) appear only on the gray dashed lines, where the generically allowed $v_z$ component is accidentally tuned to zero by specific parameter combinations. These lines are measure-zero in the parameter space, confirming that criticality at $K$ is not symmetry-enforced but rather parameter-accidental.

Taken together, these numerical results provide definitive verification of our symmetry criterion. The dichotomy between critical and noncritical 1D bands at non-TRIMs is dictated entirely by the little group's vector representation, independent of material-specific hopping amplitudes. The specific VHS subtype---whether ordinary or higher-order, and which sign pattern of Hessian eigenvalues---is parameter-dependent and requires explicit evaluation of the Hessian. This two-tier hierarchy, with symmetry governing the presence or absence of $\nabla E$ and parameters governing the higher-order character, is precisely the framework established in Sec.~\ref{sec:II}.

%%%%%%%%

%%%%%%%%%%%%%%%

\section{Extension to all space groups containing non-TRIMs}
\label{sec:IV}

Having verified the criterion in space group 225 within the single-group (spinless) limit, we now extend it systematically to all space groups containing non-TRIM high-symmetry points under the same single-group assumption.  .We have scanned all 230 space groups~\cite{bradley1972,hahn2005} and identified five crystal systems---orthorhombic, tetragonal, trigonal, hexagonal, and cubic (including both face-centered and body-centered Bravais lattices)---that contain non-TRIMs.

Before presenting the classification results in Table~\ref{tab2}, we clarify the meaning of the "Non-VHS (Excluded)" entries that appear in both branches of Fig.~\ref{fig:flowchart}. For nondegenerate bands, this exclusion occurs when $\Gamma_{\mathrm{vec}}$ contains two or three $\Gamma_1$ components, allowing multiple gradient components simultaneously. For degenerate bands, it occurs at the subband level when a particular subband has only zero or one zero component in its gradient vector after diagonalization. In both cases, the common physical consequence is linear dispersion along multiple independent directions, which yields a smooth DOS without van Hove singularities. Such points are therefore excluded from the VHS classification and typically correspond to gapless points or band crossings.

We restrict this classification to non-TRIMs for the following reason. As noted in Sec.~\ref{sec:II}, at TRIMs $\nabla E=0$ is guaranteed by time-reversal symmetry alone, rendering the application of our criterion redundant. The non-TRIMs, by contrast, lack this direct protection; their criticality depends entirely on the space-group symmetry. It is precisely at such points that our group-theoretical criterion provides a parameter-free and material-independent prediction.

The selection criterion is as follows. For each space group, we examine all high-symmetry points in its Brillouin zone. If a point is not a TRIM---i.e., there exists a space-group operation $R$ such that $R\mathbf{k}_0 \equiv \mathbf{k}_0 + \mathbf{G}$ but $-\mathbf{k}_0 \not\equiv \mathbf{k}_0 + \mathbf{G}$, where $\mathbf{G}$ is a reciprocal lattice vector---then that point is included in the classification.

For each non-TRIM point, we determine its little group and the decomposition of its vector representation $\Gamma_{\text{vec}}$. As a direct consequence of the Wigner-Eckart reduction (consistent with Neumann's principle \cite{hahn2005}), the criticality of all nondegenerate (1D) bands at a given high-symmetry point is completely universal. The classification depends strictly on the multiplicity of the trivial representation $\Gamma_1$ contained in $\Gamma_{\mathrm{vec}}$, regardless of the specific irreducible representation $\Gamma$ of the band itself:
\begin{itemize}
\item When $\Gamma_{\mathrm{vec}}$ contains \textbf{zero} $\Gamma_1$ ($\Gamma_{\mathrm{vec}} \not\supset \Gamma_1$), the gradient is strictly forced to vanish for all 1D bands, making them symmetry-enforced critical (M/T-type).
    \item When $\Gamma_{\mathrm{vec}}$ contains \textbf{exactly one} $\Gamma_1$, exactly one gradient component is allowed. All 1D bands universally exhibit one generically non-zero gradient component, rendering them noncritical (N/S-type) subject to parameter tuning (Param.-dep.).
    \item When $\Gamma_{\mathrm{vec}}$ contains \textbf{two or three} $\Gamma_1$, multiple gradient components are allowed, pushing all 1D bands into the Non-VHS (Excluded) category, typically forming gapless points or band crossings.
\end{itemize}
For degenerate bands, the analysis must proceed via their symmetrized products as outlined in Sec.~\ref{sec:II-D}. The results for both 1D and degenerate bands are summarized in Table~\ref{tab2}, where degenerate cases are marked according to whether their symmetrized products couple to one or multiple components of the gradient.

Table~\ref{tab2} presents the complete results, organized by descending little group symmetry. The classification reveals a clean pattern: little groups whose vector representation lacks the trivial representation ($T_d$, $T$, $D_{3h}$, $D_{2d}$, $C_{3h}$, $D_3$, $D_2$, $S_4$) strictly force criticality for all nondegenerate bands. Little groups whose vector representation contains exactly one trivial representation ($C_{3v}$, $C_{2v}$, $C_3$, $C_2$) universally allow exactly one non-zero gradient component, rendering all 1D bands parameter-dependent and generically noncritical. For the lowest-symmetry little groups ($C_s$ and $C_1$), the vector representation contains two or three trivial representations, respectively, and all 1D bands are excluded from the single-band VHS classification as gapless points or band crossings.

For degenerate bands (e.g., $E$ or $T$ representations), criticality depends on the intersection between their symmetrized product and $\Gamma_{\mathrm{vec}}$. For example, while the $E$ band in $D_{2d}$ couples to exactly one component ($B_2$) and is thus Param.-dep., the $E$ bands in $D_{3h}$, $D_3$, and $C_{3v}$ couple to multiple components, excluding them from the VHS classification.

From this systematic survey, we extract two design principles. First, symmetry determines whether $\nabla E$ vanishes---this is the primary distinction between critical and noncritical VHSs. Second, the specific VHS subtype (e.g., $N_0$ vs $N_1$, or $M_1$ vs $M_2$) is determined by specific band parameters and requires numerical calculations or tight-binding models for determination~\cite{shtyk2017, li2026}. These two principles constitute a two-tier framework for VHS engineering: symmetry first, parameters second.

%%%%%%%%%%%%%%%%%%%%%%%%%%%%%%
\begin{table*}[t]
\caption{Summary and classification of all space groups containing non-TRIMs in the single-group (no SOC) limit.}
\label{tab2}
\renewcommand{\arraystretch}{1.2}
\setlength{\tabcolsep}{4pt}
\centering
\scriptsize
\begin{ruledtabular}
\begin{tabular}{c c l c c c}
Little group & $\Gamma_{\text{vec}}$ & Point \& Space groups & Band IR & $\nabla E=0$? & VHS type \\
\hline
$T_d$ ($-43m$) & $T_2$ & $P$ (BCC: 217, 220, 229, 230) & $A_1 (1D), A_2 (1D), E$ (2D) & Yes & Critical ($M_0, M_3, T_3$) \\
 & & & $T_1, T_2$ (3D) & No & Excluded (Non-VHS) \\
\hline
$T$ ($23$) & $T$ & $P$ (BCC: 197, 199, 204, 206, 211, 214) & $A, {}^1E, {}^2E$ (1D) & Yes & Critical ($M_0, M_3, T_3$) \\
 & & & $T$ (3D) & No & Excluded (Non-VHS) \\
\hline
$D_{3h}$ ($-6m2$) & $A''_2\oplus E'$ & $K,H$ (Hex: 189--194) & $A'_1, A'_2, A''_1, A''_2$ (1D) & Yes & Critical (M/T) \\
 & & & $E', E''$ (2D) & No & Excluded (Non-VHS) \\
\hline
$D_{2d}$ ($-42m$) & $B_2\oplus E$ & $W$ (FCC: 225--228) & $A_1, A_2, B_1, B_2$ (1D) & Yes & Critical (M/T) \\
 & & & $E$ (2D) & No & Param.-dep. \\
\cline{3-6}
 & & $P$ (BCT: 121, 122, 139--142) & $A_1, A_2, B_1, B_2$ (1D) & Yes & Critical (M/T) \\
 & & & $E$ (2D) & No & Param.-dep. \\
\hline
$C_{3h}$ ($-6$) & $A''\oplus {}^1E' \oplus {}^2E'$ & $K,H$ (Hex: 174--176, 187, 188) & $A', A'', {}^1E', {}^2E', {}^1E'', {}^2E''$ (1D) & Yes & Critical (M/T) \\
\hline
$D_3$ ($32$) & $A_2\oplus E$ & $K,H$ (Trig: 150, 152, 154, 164, 165) & $A_1, A_2$ (1D) & Yes & Critical (M/T) \\
 & & & $E$ (2D) & No & Excluded (Non-VHS) \\
\cline{3-6}
 & & $K,H$ (Hex: 177--182) & $A_1, A_2$ (1D) & Yes & Critical (M/T) \\
 & & & $E$ (2D) & No & Excluded (Non-VHS) \\
\hline
$C_{3v}$ ($3m$) & $A_1\oplus E$ & $K,H$ (Trig: 157, 159, 162, 163) & $A_1, A_2$ (1D) & No & Param.-dep. \\
 & & & $E$ (2D) & No & Excluded (Non-VHS) \\
\cline{3-6}
 & & $K,H$ (Hex: 183--186) & $A_1, A_2$ (1D) & No & Param.-dep. \\
 & & & $E$ (2D) & No & Excluded (Non-VHS) \\
\hline
$C_{2v}$ ($mm2$) & $A_1\oplus B_1\oplus B_2$ & $W$ (FCC: 202, 203) & $A_1, A_2, B_1, B_2$ (1D) & No & Param.-dep. \\
\cline{3-6}
 & & $K,U$ (FCC: 225--228) & $A_1, A_2, B_1, B_2$ (1D) & No & Param.-dep. \\
\cline{3-6}
 & & $P$ (BCT: 107--110) & $A_1, A_2, B_1, B_2$ (1D) & No & Param.-dep. \\
\hline
$D_2$ ($222$) & $B_1\oplus B_2\oplus B_3$ & $W$ (FCC: 209, 210; BCO: 23, 24, 71--74) & $A, B_1, B_2, B_3$ (1D) & Yes & Critical (M/T) \\
\cline{3-6}
 & & $P$ (BCT: 97, 98) & $A, B_1, B_2, B_3$ (1D) & Yes & Critical (M/T) \\
\hline
$S_4$ ($-4$) & $B\oplus {}^1E \oplus {}^2E$ & $W$ (FCC: 216, 219) & $A, B, {}^1E, {}^2E$ (1D) & Yes & Critical (M/T) \\
\cline{3-6}
 & & $P$ (BCT: 82, 87, 88, 119, 120) & $A, B, {}^1E, {}^2E$ (1D) & Yes & Critical (M/T) \\
\hline
$C_3$ ($3$) & $A\oplus {}^1E \oplus {}^2E$ & $K,H$ (Trig: 143--145, 147, 149, 151, & $A, {}^1E, {}^2E$ (1D) & No & Param.-dep. \\
 & & 153, 156, 158) & & & \\
\cline{3-6}
 & & $K,H$ (Hex: 168--173) & $A, {}^1E, {}^2E$ (1D) & No & Param.-dep. \\
\hline
$C_2$ ($2$) & $A\oplus 2B$ & $W$ (FCC: 196; BCT: 44--46) & $A, B$ (1D) & No & Param.-dep. \\
\cline{3-6}
 & & $K,U$ (FCC: 209, 210) & $A, B$ (1D) & No & Param.-dep. \\
\cline{3-6}
 & & $P$ (BCT: 79, 80) & $A, B$ (1D) & No & Param.-dep. \\
\hline
$C_s$ ($m$) & $2A'\oplus A''$ & $K,U$ (FCC: 202, 203, 216, 219) & $A', A''$ (1D) & No & Excluded (Non-VHS) \\
\hline
$C_1$ ($1$) & $3A$ & $K,U$ (FCC: 196) & $A$ (1D) & No & Excluded (Non-VHS) \\
\end{tabular}
\end{ruledtabular}
\footnotetext[1]{"Yes" means $\nabla E=0$ is forced by symmetry; "No" means $\nabla E\neq 0$ is symmetry-allowed. For 1D bands, criticality follows from the multiplicity of $\Gamma_1$ in $\Gamma_{\mathrm{vec}}$: 0 $\Rightarrow$ Critical; 1 $\Rightarrow$ Param.-dep.; $\ge$2 or degenerate-band multi-component couplings $\Rightarrow$ Excluded (Non-VHS). Bands where the gradient retains only 0 or 1 zero component (which occurs when $\Gamma_{\mathrm{vec}}$ contains two or three $\Gamma_1$, or when symmetrized products of degenerate bands contain multi-component vectors) form gapless points or band crossings rather than density-of-states singularities, and are therefore marked as ``Excluded (Non-VHS)''.}
\footnotetext[2]{The notations ${}^1E,{}^2E$ (and ${}^1E',{}^2E'$, etc.) denote 1D complex-conjugate representations. Though time-reversal symmetry often pairs them into two-dimensional (2D) Kramers-degenerate bands, our criterion treats them individually to expose phase independence; both in each pair yield identical criticality.}
\end{table*}

\section{Discussion and Conclusion}

This work has revealed the symmetry origin of the dichotomy between critical and noncritical Van Hove singularities through a general group-theoretical criterion for predicting the criticality of electronic bands at non-TRIM high-symmetry points in the single-group limit. For all nondegenerate bands, the Wigner-Eckart theorem reduces the criterion to a single property of the little group: the multiplicity of the trivial representation $\Gamma_1$ contained in the vector representation $\Gamma_{\mathrm{vec}}$. Multiplicity zero forces $\nabla E=0$ (symmetry-enforced critical); multiplicity one allows exactly one nonzero gradient component (generically noncritical, with the actual magnitude determined by band parameters); multiplicity two or three excludes the band from VHS classification. For degenerate bands, the symmetrized product $[\Gamma \otimes \Gamma]_{\mathrm{sym}}$ must be analyzed via Clebsch--Gordan coefficients, and the classification is performed at the subband level.

Applying this criterion to space group 225, we have explained the profound universality within its high-symmetry points. The $W$ point ($D_{2d}$, $\Gamma_{\mathrm{vec}} = B_2 \oplus E$, lacking $A_1$) is universally critical for all nondegenerate bands. In stark contrast, at the $K$ and $U$ points ($C_{2v}$, $\Gamma_{\mathrm{vec}} = A_1 \oplus B_1 \oplus B_2$, containing exactly one $A_1$), all one-dimensional bands generically possess exactly one nonzero gradient component, rendering them noncritical. These symmetry predictions are completely verified by the tight-binding phase diagrams in Fig.~\ref{fig:phasediagram}. Beyond confirming the critical versus noncritical dichotomy, the phase diagrams reveal a clear hierarchy: symmetry enforces the vanishing of $\nabla E$ throughout finite regions of parameter space, while higher-order features (T-type and S-type) requires additional parameter tuning and appears only on lines or points where the Hessian becomes singular. This hierarchy---symmetry first, parameters second---is a defining feature of our framework.

We have systematically extended the criterion to all space groups containing non-TRIMs, covering orthorhombic, tetragonal, trigonal, hexagonal, and cubic systems, and presented a complete classification in Table~\ref{tab2}. The survey reveals a clean pattern: little groups whose vector representation lacks $\Gamma_1$ ($T_d, T, D_{3h}, D_{2d}, C_{3h}, D_3, D_2, S_4$) strictly force criticality for all nondegenerate bands. Little groups whose vector representation contains exactly one $\Gamma_1$ ($C_{3v}, C_{2v}, C_3, C_2$) universally allow exactly one nonzero gradient component, rendering all one-dimensional bands parameter-dependent and generically noncritical. Little groups with two or three $\Gamma_1$ components ($C_s, C_1$) exclude all one-dimensional bands from the VHS classification, as multiple gradient components are allowed, typically forming gapless points or band crossings rather than DOS singularities.

From this systematic survey, we extract two design principles for engineering VHSs in quantum materials. First, symmetry---specifically, the multiplicity of $\Gamma_1$ in $\Gamma_{\mathrm{vec}}$---determines whether $\nabla E$ vanishes, providing the fundamental distinction between critical and noncritical VHSs. Second, the specific VHS subtype (e.g., $M_0$ vs $M_1$, or $N_0$ vs $N_1$) is determined by band parameters and requires numerical evaluation of hopping amplitudes. This two-tier structure provides a practical, parameter-free diagnostic tool: consulting Table~\ref{tab2} immediately reveals whether a given non-TRIM point is symmetry-enforced critical or generically noncritical, without requiring any band-structure calculation. Only the specific subtype classification requires explicit tight-binding or first-principles evaluation. This capability makes our framework directly applicable to high-throughput materials screening and experimental band-structure engineering.

Finally, we emphasize that the conclusions of this work are firmly rooted in the single-particle band theory of weakly correlated, paramagnetic systems in the single-group (no SOC) limit. For systems with strong spin-orbit coupling or magnetic ordering, the present single-group classification does not directly apply; however, the framework can be systematically adapted using double-group or magnetic little-group representations, respectively, which we leave for future work. Our work thus not only resolves the symmetry origin of criticality in conventional Van Hove singularities but also lays a solid foundation for extending such symmetry-based classifications to more exotic symmetry-broken and relativistic regimes.

Together with our previous work~\cite{li2026}---which established the unified classification of VHS types and demonstrated their numerical realization in the pyrochlore lattice---this work provides the rigorous group-theoretical foundation and extends the framework to all 230 space groups. This constitutes a complete theoretical framework for the structural engineering of electronic densities of states in three-dimensional quantum materials.

\section*{Acknowledgments}
M.Q. Kuang acknowledges the support from the Natural Science Foundation of Chongqing (Grant No. CSTB2024NSCQ-MSX0080) and the National Natural Science Foundation of China (NSFC, Grant No. 11704315).

%%%%%%%%%%%%%% Appdeix
\begin{appendices}

\section{Tight-Binding Model and Phase Diagram Calculation}
\label{sec:TB225}
To mathematically verify the group-theoretical criteria and explicitly demonstrate how the phase diagrams in Fig.~\ref{fig:phasediagram} are constructed, we present the exact analytical tight-binding Hamiltonian for space group 225 on the 4$a$ Wyckoff position $(0,0,0)$, generated using the MagneticTB package~\cite{zhang2022}. 

We construct the model using $s$-orbital basis functions. In the presence of SOC, the Hamiltonian would be a $2 \times 2$ block-diagonal matrix $\mathcal{H}_{\text{SOC}} = \text{diag}(A, A)$. However, because we are investigating the single-group limit (explicitly excluding SOC), the Hamiltonian simplifies to a $1 \times 1$ scalar matrix $\mathcal{H}_0(\mathbf{k}) = (A)$. The single matrix element $A$ directly serves as the analytical energy band expression $E(\mathbf{k})$ for the studied 1D band.

Incorporating the on-site energy $e_1$, nearest-neighbor ($t_1$), second-neighbor ($r_1$), and third-neighbor ($s_1$) hopping amplitudes, the exact band expression evaluates to:
\begin{align}
E(\mathbf{k}) =\;& e_1 + t_1 \sum_{\mathbf{d}_1} e^{i \mathbf{k} \cdot \mathbf{d}_1} + r_1 \sum_{\mathbf{d}_2} e^{i \mathbf{k} \cdot \mathbf{d}_2} \nonumber + s_1 \sum_{\mathbf{d}_3} e^{i \mathbf{k} \cdot \mathbf{d}_3} \nonumber \\
=\;& e_1 + 2t_1 \big[ \cos(k_x) + \cos(k_y) + \cos(k_z) \nonumber \\
& + \cos(k_x-k_y) + \cos(k_y-k_z) + \cos(k_x-k_z) \big] \nonumber \\
& + 2r_1 \big[ \cos(k_x-k_y-k_z) + \cos(k_x-k_y+k_z) \nonumber \\
& + \cos(k_x+k_y-k_z) \big] \nonumber \\
& + 2s_1 \big[ \cos(k_x+k_y) + \cos(k_y+k_z) \nonumber \\
& + \cos(k_x+k_z) + \cos(2k_x-k_y) + \cos(k_x-2k_y) \nonumber \\
& + \cos(2k_x-k_z) + \cos(k_x-2k_z) + \cos(2k_y-k_z) \nonumber \\
& + \cos(k_y-2k_z) + \cos(2k_x-k_y-k_z) \nonumber \\
& + \cos(k_x-2k_y+k_z) + \cos(k_x+k_y-2k_z) \big]
\label{eq:band_expr}
\end{align}
where the summations run over all symmetric bond vectors for each nearest-neighbor shell derived from the FCC lattice geometry.

To mathematically determine the VHS type at a specific high-symmetry point $\mathbf{k}_0$ (such as $W$ or $K$), we perform a Taylor expansion of the band expression Eq.~(\ref{eq:band_expr}) around $\mathbf{k}_0$ by setting $\mathbf{k} = \mathbf{k}_0 + \mathbf{q}$. This expansion is equivalent to Eq.~\eqref{eq:expansion} in matrix-vector notation:
\begin{equation}
E(\mathbf{k}_0 + \mathbf{q}) = E(\mathbf{k}_0) + \mathbf{v} \cdot \mathbf{q} + \frac{1}{2} \mathbf{q}^T \mathcal{H} \mathbf{q} + \mathcal{O}(q^3).
\end{equation}
Here, $\mathbf{v} = \nabla_{\mathbf{k}} E|_{\mathbf{k}_0}$ is the gradient vector, and $H_{ij} = \partial^2 E/\partial k_i \partial k_j|_{\mathbf{k}_0}$ is the $3 \times 3$ Hessian matrix.

The classification of the VHS subtype proceeds algorithmically based on these analytic derivatives:
\begin{enumerate}
    \item \textbf{Gradient evaluation (Critical vs. Noncritical):} We first calculate the vector $\mathbf{v}$. If $|\mathbf{v}| \neq 0$, the singularity is classified as noncritical ($N$- or $S$-type). If $\mathbf{v} = \mathbf{0}$ (either enforced globally by symmetry, as at the $W$ point, or through accidental parameter tuning, as on the dashed lines of the K point), the singularity is critical ($M$- or $T$-type).
    \item \textbf{Hessian eigenvalue analysis:} We then diagonalize the Hessian matrix $\mathcal{H}$ to obtain its three eigenvalues $\lambda_1, \lambda_2, \lambda_3$. 
    \item \textbf{Subtype assignment:} For a critical VHS ($\mathbf{v} = \mathbf{0}$), if all $\lambda_i \neq 0$, it is an ordinary critical point ($M_0$ or $M_3$ if signs are identical; $M_1$ or $M_2$ if signs are mixed). If one or more $\lambda_i = 0$ (i.e., the Hessian determinant vanishes, $\det \mathcal{H} = 0$), the dispersion is locally flat, yielding a higher-order critical point ($T_1, T_2, T_3$). For a noncritical VHS ($\mathbf{v} \neq \mathbf{0}$), a similar eigenvalue analysis restricted to the two-dimensional subspace orthogonal to $\mathbf{v}$ determines whether it is an ordinary ($N_0, N_1, N_2$) or higher-order ($S_1, S_2$) noncritical point~\cite{li2026}.
\end{enumerate}

By setting $t_1 = -1$ as the energy unit and scanning the parameters $r_1$ and $s_1$ over the range $[-1, 1]$, this strict mathematical procedure is applied at every coordinate $(r_1, s_1)$ to construct the exact phase boundaries presented in Fig.~\ref{fig:phasediagram}.

\end{appendices}
%%%%%%%%%%%%%%%%
\bibliography{refs}

\begin{thebibliography}{41}%
\makeatletter
\providecommand \@ifxundefined [1]{%
 \@ifx{#1\undefined}
}%
\providecommand \@ifnum [1]{%
 \ifnum #1\expandafter \@firstoftwo
 \else \expandafter \@secondoftwo
 \fi
}%
\providecommand \@ifx [1]{%
 \ifx #1\expandafter \@firstoftwo
 \else \expandafter \@secondoftwo
 \fi
}%
\providecommand \natexlab [1]{#1}%
\providecommand \enquote  [1]{``#1''}%
\providecommand \bibnamefont  [1]{#1}%
\providecommand \bibfnamefont [1]{#1}%
\providecommand \citenamefont [1]{#1}%
\providecommand \href@noop [0]{\@secondoftwo}%
\providecommand \href [0]{\begingroup \@sanitize@url \@href}%
\providecommand \@href[1]{\@@startlink{#1}\@@href}%
\providecommand \@@href[1]{\endgroup#1\@@endlink}%
\providecommand \@sanitize@url [0]{\catcode `\\12\catcode `\$12\catcode `\&12\catcode `\#12\catcode `\^12\catcode `\_12\catcode `\%12\relax}%
\providecommand \@@startlink[1]{}%
\providecommand \@@endlink[0]{}%
\providecommand \url  [0]{\begingroup\@sanitize@url \@url }%
\providecommand \@url [1]{\endgroup\@href {#1}{\urlprefix }}%
\providecommand \urlprefix  [0]{URL }%
\providecommand \Eprint [0]{\href }%
\providecommand \doibase [0]{https://doi.org/}%
\providecommand \selectlanguage [0]{\@gobble}%
\providecommand \bibinfo  [0]{\@secondoftwo}%
\providecommand \bibfield  [0]{\@secondoftwo}%
\providecommand \translation [1]{[#1]}%
\providecommand \BibitemOpen [0]{}%
\providecommand \bibitemStop [0]{}%
\providecommand \bibitemNoStop [0]{.\EOS\space}%
\providecommand \EOS [0]{\spacefactor3000\relax}%
\providecommand \BibitemShut  [1]{\csname bibitem#1\endcsname}%
\let\auto@bib@innerbib\@empty
%</preamble>
\bibitem [{\citenamefont {Van~Hove}(1953)}]{vanhove1953}%
  \BibitemOpen
  \bibfield  {author} {\bibinfo {author} {\bibfnamefont {L.}~\bibnamefont {Van~Hove}},\ }\bibfield  {title} {\bibinfo {title} {The occurrence of singularities in the elastic frequency distribution of a crystal},\ }\href {https://doi.org/10.1103/PhysRev.89.1189} {\bibfield  {journal} {\bibinfo  {journal} {Physical Review}\ }\textbf {\bibinfo {volume} {89}},\ \bibinfo {pages} {1189} (\bibinfo {year} {1953})}\BibitemShut {NoStop}%
\bibitem [{\citenamefont {Efremov}\ \emph {et~al.}(2019)\citenamefont {Efremov}, \citenamefont {Shtyk}, \citenamefont {Rost}, \citenamefont {Chamon}, \citenamefont {Mackenzie},\ and\ \citenamefont {Betouras}}]{efremov2019}%
  \BibitemOpen
  \bibfield  {author} {\bibinfo {author} {\bibfnamefont {D.~V.}\ \bibnamefont {Efremov}}, \bibinfo {author} {\bibfnamefont {A.}~\bibnamefont {Shtyk}}, \bibinfo {author} {\bibfnamefont {A.~W.}\ \bibnamefont {Rost}}, \bibinfo {author} {\bibfnamefont {C.}~\bibnamefont {Chamon}}, \bibinfo {author} {\bibfnamefont {A.~P.}\ \bibnamefont {Mackenzie}},\ and\ \bibinfo {author} {\bibfnamefont {J.~J.}\ \bibnamefont {Betouras}},\ }\bibfield  {title} {\bibinfo {title} {Multicritical {F}ermi surface topological transitions},\ }\href {https://doi.org/10.1103/PhysRevLett.123.207202} {\bibfield  {journal} {\bibinfo  {journal} {Physical Review Letters}\ }\textbf {\bibinfo {volume} {123}},\ \bibinfo {pages} {207202} (\bibinfo {year} {2019})}\BibitemShut {NoStop}%
\bibitem [{\citenamefont {Yuan}\ \emph {et~al.}(2019)\citenamefont {Yuan}, \citenamefont {Isobe},\ and\ \citenamefont {Fu}}]{yuan2019}%
  \BibitemOpen
  \bibfield  {author} {\bibinfo {author} {\bibfnamefont {N.~F.}\ \bibnamefont {Yuan}}, \bibinfo {author} {\bibfnamefont {H.}~\bibnamefont {Isobe}},\ and\ \bibinfo {author} {\bibfnamefont {L.}~\bibnamefont {Fu}},\ }\bibfield  {title} {\bibinfo {title} {Magic of high-order van {H}ove singularity},\ }\href {https://doi.org/10.1038/s41467-019-13670-9} {\bibfield  {journal} {\bibinfo  {journal} {Nat. Commun}\ }\textbf {\bibinfo {volume} {10}},\ \bibinfo {pages} {5769} (\bibinfo {year} {2019})}\BibitemShut {NoStop}%
\bibitem [{\citenamefont {Yuan}\ and\ \citenamefont {Fu}(2020)}]{yuan2020}%
  \BibitemOpen
  \bibfield  {author} {\bibinfo {author} {\bibfnamefont {N.~F.~Q.}\ \bibnamefont {Yuan}}\ and\ \bibinfo {author} {\bibfnamefont {L.}~\bibnamefont {Fu}},\ }\bibfield  {title} {\bibinfo {title} {Classification of critical points in energy bands based on topology, scaling, and symmetry},\ }\href {https://doi.org/10.1103/PhysRevB.101.125120} {\bibfield  {journal} {\bibinfo  {journal} {Physical Review B}\ }\textbf {\bibinfo {volume} {101}},\ \bibinfo {pages} {125120} (\bibinfo {year} {2020})}\BibitemShut {NoStop}%
\bibitem [{\citenamefont {Patra}\ \emph {et~al.}(2025)\citenamefont {Patra}, \citenamefont {Mukherjee},\ and\ \citenamefont {Singh}}]{patra2025}%
  \BibitemOpen
  \bibfield  {author} {\bibinfo {author} {\bibfnamefont {B.}~\bibnamefont {Patra}}, \bibinfo {author} {\bibfnamefont {A.}~\bibnamefont {Mukherjee}},\ and\ \bibinfo {author} {\bibfnamefont {B.}~\bibnamefont {Singh}},\ }\bibfield  {title} {\bibinfo {title} {High-order van {H}ove singularities and nematic instability in the kagome superconductor {CsTi$_3$Bi$_5$}},\ }\href {https://doi.org/10.1103/PhysRevB.111.045135} {\bibfield  {journal} {\bibinfo  {journal} {Physical Review B}\ }\textbf {\bibinfo {volume} {111}},\ \bibinfo {pages} {045135} (\bibinfo {year} {2025})}\BibitemShut {NoStop}%
\bibitem [{\citenamefont {Classen}\ and\ \citenamefont {Betouras}(2025)}]{classen2025}%
  \BibitemOpen
  \bibfield  {author} {\bibinfo {author} {\bibfnamefont {L.}~\bibnamefont {Classen}}\ and\ \bibinfo {author} {\bibfnamefont {J.~J.}\ \bibnamefont {Betouras}},\ }\bibfield  {title} {\bibinfo {title} {High-order van {H}ove singularities and their connection to flat bands},\ }\href {https://doi.org/10.1146/annurev-conmatphys-042924-015000} {\bibfield  {journal} {\bibinfo  {journal} {Annual Review of Condensed Matter Physics}\ }\textbf {\bibinfo {volume} {16}},\ \bibinfo {pages} {229} (\bibinfo {year} {2025})}\BibitemShut {NoStop}%
\bibitem [{\citenamefont {Lifshitz}(1960)}]{lifshitz1960}%
  \BibitemOpen
  \bibfield  {author} {\bibinfo {author} {\bibfnamefont {I.~M.}\ \bibnamefont {Lifshitz}},\ }\bibfield  {title} {\bibinfo {title} {Anomalies of electron characteristics of a metal in the high pressure region},\ }\href@noop {} {\bibfield  {journal} {\bibinfo  {journal} {Soviet Physics JETP}\ }\textbf {\bibinfo {volume} {11}},\ \bibinfo {pages} {1130} (\bibinfo {year} {1960})},\ \bibinfo {note} {translated from Zh. Eksp. Teor. Fiz. 38, 156 (1960)}\BibitemShut {NoStop}%
\bibitem [{\citenamefont {Tamai}\ \emph {et~al.}(2008)\citenamefont {Tamai}, \citenamefont {Allan}, \citenamefont {Mercure}, \citenamefont {Meevasana}, \citenamefont {Dunkel}, \citenamefont {Lu}, \citenamefont {Perry}, \citenamefont {Mackenzie}, \citenamefont {Singh},\ and\ \citenamefont {Shen}}]{tamai2008}%
  \BibitemOpen
  \bibfield  {author} {\bibinfo {author} {\bibfnamefont {A.}~\bibnamefont {Tamai}}, \bibinfo {author} {\bibfnamefont {M.~P.}\ \bibnamefont {Allan}}, \bibinfo {author} {\bibfnamefont {J.-F.}\ \bibnamefont {Mercure}}, \bibinfo {author} {\bibfnamefont {W.}~\bibnamefont {Meevasana}}, \bibinfo {author} {\bibfnamefont {R.}~\bibnamefont {Dunkel}}, \bibinfo {author} {\bibfnamefont {D.}~\bibnamefont {Lu}}, \bibinfo {author} {\bibfnamefont {R.~S.}\ \bibnamefont {Perry}}, \bibinfo {author} {\bibfnamefont {A.~P.}\ \bibnamefont {Mackenzie}}, \bibinfo {author} {\bibfnamefont {D.~J.}\ \bibnamefont {Singh}},\ and\ \bibinfo {author} {\bibfnamefont {Z.-X.}\ \bibnamefont {Shen}},\ }\bibfield  {title} {\bibinfo {title} {Fermi surface and van {H}ove singularities in the itinerant metamagnet {Sr$_3$Ru$_2$O$_7$}},\ }\href {https://doi.org/10.1103/PhysRevLett.101.026407} {\bibfield  {journal} {\bibinfo  {journal} {Phys. Rev. Lett.}\ }\textbf {\bibinfo {volume} {101}},\ \bibinfo {pages} {026407} (\bibinfo {year} {2008})}\BibitemShut
  {NoStop}%
\bibitem [{\citenamefont {Wu}\ \emph {et~al.}(2021)\citenamefont {Wu}, \citenamefont {Schwemmer}, \citenamefont {M{\"u}ller}, \citenamefont {Consiglio}, \citenamefont {Sangiovanni}, \citenamefont {Di~Sante}, \citenamefont {Iqbal}, \citenamefont {Hanke}, \citenamefont {Schnyder}, \citenamefont {Denner}, \citenamefont {Neupert},\ and\ \citenamefont {Thomale}}]{wu2021}%
  \BibitemOpen
  \bibfield  {author} {\bibinfo {author} {\bibfnamefont {X.}~\bibnamefont {Wu}}, \bibinfo {author} {\bibfnamefont {T.}~\bibnamefont {Schwemmer}}, \bibinfo {author} {\bibfnamefont {T.}~\bibnamefont {M{\"u}ller}}, \bibinfo {author} {\bibfnamefont {A.}~\bibnamefont {Consiglio}}, \bibinfo {author} {\bibfnamefont {G.}~\bibnamefont {Sangiovanni}}, \bibinfo {author} {\bibfnamefont {D.}~\bibnamefont {Di~Sante}}, \bibinfo {author} {\bibfnamefont {Y.}~\bibnamefont {Iqbal}}, \bibinfo {author} {\bibfnamefont {W.}~\bibnamefont {Hanke}}, \bibinfo {author} {\bibfnamefont {A.~P.}\ \bibnamefont {Schnyder}}, \bibinfo {author} {\bibfnamefont {M.~M.}\ \bibnamefont {Denner}}, \bibinfo {author} {\bibfnamefont {T.}~\bibnamefont {Neupert}},\ and\ \bibinfo {author} {\bibfnamefont {R.}~\bibnamefont {Thomale}},\ }\bibfield  {title} {\bibinfo {title} {Nature of unconventional pairing in the kagome superconductors {$A$V$_3$Sb$_5$ ($A = $ K, Rb, Cs)}},\ }\href {https://doi.org/10.1103/PhysRevLett.127.177001} {\bibfield  {journal}
  {\bibinfo  {journal} {Phys. Rev. Lett.}\ }\textbf {\bibinfo {volume} {127}},\ \bibinfo {pages} {177001} (\bibinfo {year} {2021})}\BibitemShut {NoStop}%
\bibitem [{\citenamefont {Tan}\ \emph {et~al.}(2024)\citenamefont {Tan}, \citenamefont {Jiang}, \citenamefont {McCandless}, \citenamefont {Chan},\ and\ \citenamefont {Yan}}]{tan2024}%
  \BibitemOpen
  \bibfield  {author} {\bibinfo {author} {\bibfnamefont {H.}~\bibnamefont {Tan}}, \bibinfo {author} {\bibfnamefont {Y.}~\bibnamefont {Jiang}}, \bibinfo {author} {\bibfnamefont {G.~T.}\ \bibnamefont {McCandless}}, \bibinfo {author} {\bibfnamefont {J.~Y.}\ \bibnamefont {Chan}},\ and\ \bibinfo {author} {\bibfnamefont {B.}~\bibnamefont {Yan}},\ }\bibfield  {title} {\bibinfo {title} {Three-dimensional higher-order saddle-point-induced flat bands in {C}o-based kagome metals},\ }\href {https://doi.org/10.1103/PhysRevResearch.6.043132} {\bibfield  {journal} {\bibinfo  {journal} {Phys. Rev. Res.}\ }\textbf {\bibinfo {volume} {6}},\ \bibinfo {pages} {043132} (\bibinfo {year} {2024})}\BibitemShut {NoStop}%
\bibitem [{\citenamefont {Li}\ \emph {et~al.}(2026)\citenamefont {Li}, \citenamefont {Tan}, \citenamefont {Zhu}, \citenamefont {Yuan},\ and\ \citenamefont {Kuang}}]{li2026}%
  \BibitemOpen
  \bibfield  {author} {\bibinfo {author} {\bibfnamefont {H.-Y.}\ \bibnamefont {Li}}, \bibinfo {author} {\bibfnamefont {H.}~\bibnamefont {Tan}}, \bibinfo {author} {\bibfnamefont {H.-Y.}\ \bibnamefont {Zhu}}, \bibinfo {author} {\bibfnamefont {H.-K.}\ \bibnamefont {Yuan}},\ and\ \bibinfo {author} {\bibfnamefont {M.-Q.}\ \bibnamefont {Kuang}},\ }\bibfield  {title} {\bibinfo {title} {Directional criticality and higher-order flatness: Designing van {H}ove singularities in three dimensions},\ }\bibfield  {journal} {\bibinfo  {journal} {arXiv preprint}\ }\href {https://doi.org/10.48550/arXiv.2604.07806} {10.48550/arXiv.2604.07806} (\bibinfo {year} {2026}),\ \Eprint {https://arxiv.org/abs/2604.07806} {arXiv:2604.07806} \BibitemShut {NoStop}%
\bibitem [{\citenamefont {Liu}\ \emph {et~al.}(2013)\citenamefont {Liu}, \citenamefont {Duan},\ and\ \citenamefont {Fu}}]{liu2013}%
  \BibitemOpen
  \bibfield  {author} {\bibinfo {author} {\bibfnamefont {J.}~\bibnamefont {Liu}}, \bibinfo {author} {\bibfnamefont {W.}~\bibnamefont {Duan}},\ and\ \bibinfo {author} {\bibfnamefont {L.}~\bibnamefont {Fu}},\ }\bibfield  {title} {\bibinfo {title} {Two types of surface states in topological crystalline insulators},\ }\href {https://doi.org/10.1103/PhysRevB.88.241303} {\bibfield  {journal} {\bibinfo  {journal} {Physical Review B}\ }\textbf {\bibinfo {volume} {88}},\ \bibinfo {pages} {241303(R)} (\bibinfo {year} {2013})}\BibitemShut {NoStop}%
\bibitem [{\citenamefont {Luo}\ \emph {et~al.}(2025)\citenamefont {Luo}, \citenamefont {Pan}, \citenamefont {Shi},\ and\ \citenamefont {Wu}}]{luo2025}%
  \BibitemOpen
  \bibfield  {author} {\bibinfo {author} {\bibfnamefont {X.-J.}\ \bibnamefont {Luo}}, \bibinfo {author} {\bibfnamefont {X.-H.}\ \bibnamefont {Pan}}, \bibinfo {author} {\bibfnamefont {Y.}~\bibnamefont {Shi}},\ and\ \bibinfo {author} {\bibfnamefont {F.}~\bibnamefont {Wu}},\ }\bibfield  {title} {\bibinfo {title} {Surface-dependent {M}ajorana vortex phases in topological crystalline insulators},\ }\href {https://doi.org/10.1103/PhysRevB.111.144501} {\bibfield  {journal} {\bibinfo  {journal} {Physical Review B}\ }\textbf {\bibinfo {volume} {111}},\ \bibinfo {pages} {144501} (\bibinfo {year} {2025})}\BibitemShut {NoStop}%
\bibitem [{\citenamefont {Sarmah}\ \emph {et~al.}(2025)\citenamefont {Sarmah}, \citenamefont {Dutta}, \citenamefont {Ghosh},\ and\ \citenamefont {Dasgupta}}]{sarmah2025}%
  \BibitemOpen
  \bibfield  {author} {\bibinfo {author} {\bibfnamefont {H.~S.}\ \bibnamefont {Sarmah}}, \bibinfo {author} {\bibfnamefont {K.}~\bibnamefont {Dutta}}, \bibinfo {author} {\bibfnamefont {S.}~\bibnamefont {Ghosh}},\ and\ \bibinfo {author} {\bibfnamefont {I.}~\bibnamefont {Dasgupta}},\ }\bibfield  {title} {\bibinfo {title} {Rashba and {Z}eeman splitting in non-magnetic and non-centrosymmetric {MXene} {Ta$_2$CS$_2$}},\ }\href {https://doi.org/https://doi.org/10.1103/pph8-2839} {\bibfield  {journal} {\bibinfo  {journal} {Physical Review Materials}\ }\textbf {\bibinfo {volume} {9}},\ \bibinfo {pages} {074004} (\bibinfo {year} {2025})}\BibitemShut {NoStop}%
\bibitem [{\citenamefont {Tao}\ \emph {et~al.}(2023)\citenamefont {Tao}, \citenamefont {Li}, \citenamefont {Liu}, \citenamefont {Wang}, \citenamefont {Sui}, \citenamefont {Song}, \citenamefont {Zhuravlev},\ and\ \citenamefont {Liu}}]{tao2023}%
  \BibitemOpen
  \bibfield  {author} {\bibinfo {author} {\bibfnamefont {L.}~\bibnamefont {Tao}}, \bibinfo {author} {\bibfnamefont {J.}~\bibnamefont {Li}}, \bibinfo {author} {\bibfnamefont {Y.}~\bibnamefont {Liu}}, \bibinfo {author} {\bibfnamefont {X.}~\bibnamefont {Wang}}, \bibinfo {author} {\bibfnamefont {Y.}~\bibnamefont {Sui}}, \bibinfo {author} {\bibfnamefont {B.}~\bibnamefont {Song}}, \bibinfo {author} {\bibfnamefont {M.~Y.}\ \bibnamefont {Zhuravlev}},\ and\ \bibinfo {author} {\bibfnamefont {Q.}~\bibnamefont {Liu}},\ }\bibfield  {title} {\bibinfo {title} {Rashba spin splitting around non-time-reversal-invariant momenta},\ }\href {https://doi.org/10.1103/PhysRevB.107.235138} {\bibfield  {journal} {\bibinfo  {journal} {Physical Review B}\ }\textbf {\bibinfo {volume} {107}},\ \bibinfo {pages} {235138} (\bibinfo {year} {2023})}\BibitemShut {NoStop}%
\bibitem [{\citenamefont {Zhu}\ \emph {et~al.}(2022)\citenamefont {Zhu}, \citenamefont {Wu}, \citenamefont {Zhao}, \citenamefont {Chen}, \citenamefont {Zhang},\ and\ \citenamefont {Yang}}]{zhu2022}%
  \BibitemOpen
  \bibfield  {author} {\bibinfo {author} {\bibfnamefont {J.}~\bibnamefont {Zhu}}, \bibinfo {author} {\bibfnamefont {W.}~\bibnamefont {Wu}}, \bibinfo {author} {\bibfnamefont {J.}~\bibnamefont {Zhao}}, \bibinfo {author} {\bibfnamefont {H.}~\bibnamefont {Chen}}, \bibinfo {author} {\bibfnamefont {L.}~\bibnamefont {Zhang}},\ and\ \bibinfo {author} {\bibfnamefont {S.~A.}\ \bibnamefont {Yang}},\ }\bibfield  {title} {\bibinfo {title} {Symmetry-enforced nodal chain phonons},\ }\href {https://doi.org/10.1038/s41535-022-00461-7} {\bibfield  {journal} {\bibinfo  {journal} {npj Quantum Materials}\ }\textbf {\bibinfo {volume} {7}},\ \bibinfo {pages} {52} (\bibinfo {year} {2022})}\BibitemShut {NoStop}%
\bibitem [{\citenamefont {Bradley}\ and\ \citenamefont {Cracknell}(1972)}]{bradley1972}%
  \BibitemOpen
  \bibfield  {author} {\bibinfo {author} {\bibfnamefont {C.~J.}\ \bibnamefont {Bradley}}\ and\ \bibinfo {author} {\bibfnamefont {A.~P.}\ \bibnamefont {Cracknell}},\ }\href@noop {} {\emph {\bibinfo {title} {The Mathematical Theory of Symmetry in Solids}}},\ Oxford Mathematical Monographs\ (\bibinfo  {publisher} {Clarendon Press},\ \bibinfo {address} {Oxford},\ \bibinfo {year} {1972})\ \bibinfo {note} {reprinted in the Oxford Classic Texts in the Physical Sciences series}\BibitemShut {NoStop}%
\bibitem [{\citenamefont {Herring}(1937)}]{herring1937}%
  \BibitemOpen
  \bibfield  {author} {\bibinfo {author} {\bibfnamefont {C.}~\bibnamefont {Herring}},\ }\bibfield  {title} {\bibinfo {title} {Effect of time-reversal symmetry on the energy bands of crystals},\ }\href {https://doi.org/10.1103/PhysRev.52.361} {\bibfield  {journal} {\bibinfo  {journal} {Phys. Rev.}\ }\textbf {\bibinfo {volume} {52}},\ \bibinfo {pages} {361} (\bibinfo {year} {1937})}\BibitemShut {NoStop}%
\bibitem [{\citenamefont {Bouckaert}\ \emph {et~al.}(1936)\citenamefont {Bouckaert}, \citenamefont {Smoluchowski},\ and\ \citenamefont {Wigner}}]{bouckaert1936}%
  \BibitemOpen
  \bibfield  {author} {\bibinfo {author} {\bibfnamefont {L.~P.}\ \bibnamefont {Bouckaert}}, \bibinfo {author} {\bibfnamefont {R.}~\bibnamefont {Smoluchowski}},\ and\ \bibinfo {author} {\bibfnamefont {E.}~\bibnamefont {Wigner}},\ }\bibfield  {title} {\bibinfo {title} {Theory of {B}rillouin zones and symmetry properties of wave functions in crystals},\ }\href {https://doi.org/10.1103/PhysRev.50.58} {\bibfield  {journal} {\bibinfo  {journal} {Phys. Rev.}\ }\textbf {\bibinfo {volume} {50}},\ \bibinfo {pages} {58} (\bibinfo {year} {1936})}\BibitemShut {NoStop}%
\bibitem [{\citenamefont {Weng}\ \emph {et~al.}(2015)\citenamefont {Weng}, \citenamefont {Fang}, \citenamefont {Fang}, \citenamefont {Bernevig},\ and\ \citenamefont {Dai}}]{weng2015}%
  \BibitemOpen
  \bibfield  {author} {\bibinfo {author} {\bibfnamefont {H.}~\bibnamefont {Weng}}, \bibinfo {author} {\bibfnamefont {C.}~\bibnamefont {Fang}}, \bibinfo {author} {\bibfnamefont {Z.}~\bibnamefont {Fang}}, \bibinfo {author} {\bibfnamefont {B.~A.}\ \bibnamefont {Bernevig}},\ and\ \bibinfo {author} {\bibfnamefont {X.}~\bibnamefont {Dai}},\ }\bibfield  {title} {\bibinfo {title} {Weyl semimetal phase in noncentrosymmetric transition-metal monophosphides},\ }\href {https://doi.org/10.1103/PhysRevX.5.011029} {\bibfield  {journal} {\bibinfo  {journal} {Phys. Rev. X}\ }\textbf {\bibinfo {volume} {5}},\ \bibinfo {pages} {011029} (\bibinfo {year} {2015})}\BibitemShut {NoStop}%
\bibitem [{\citenamefont {Bradlyn}\ \emph {et~al.}(2016)\citenamefont {Bradlyn}, \citenamefont {Cano}, \citenamefont {Wang}, \citenamefont {Vergniory}, \citenamefont {Felser}, \citenamefont {Cava},\ and\ \citenamefont {Bernevig}}]{bradlyn2016}%
  \BibitemOpen
  \bibfield  {author} {\bibinfo {author} {\bibfnamefont {B.}~\bibnamefont {Bradlyn}}, \bibinfo {author} {\bibfnamefont {J.}~\bibnamefont {Cano}}, \bibinfo {author} {\bibfnamefont {Z.}~\bibnamefont {Wang}}, \bibinfo {author} {\bibfnamefont {M.~G.}\ \bibnamefont {Vergniory}}, \bibinfo {author} {\bibfnamefont {C.}~\bibnamefont {Felser}}, \bibinfo {author} {\bibfnamefont {R.~J.}\ \bibnamefont {Cava}},\ and\ \bibinfo {author} {\bibfnamefont {B.~A.}\ \bibnamefont {Bernevig}},\ }\bibfield  {title} {\bibinfo {title} {Beyond {D}irac and {W}eyl fermions: Unconventional quasiparticles in conventional crystals},\ }\href {https://doi.org/10.1126/science.aaf5037} {\bibfield  {journal} {\bibinfo  {journal} {Science}\ }\textbf {\bibinfo {volume} {353}},\ \bibinfo {pages} {aaf5037} (\bibinfo {year} {2016})}\BibitemShut {NoStop}%
\bibitem [{\citenamefont {Armitage}\ \emph {et~al.}(2018)\citenamefont {Armitage}, \citenamefont {Mele},\ and\ \citenamefont {Vishwanath}}]{armitage2018}%
  \BibitemOpen
  \bibfield  {author} {\bibinfo {author} {\bibfnamefont {N.~P.}\ \bibnamefont {Armitage}}, \bibinfo {author} {\bibfnamefont {E.~J.}\ \bibnamefont {Mele}},\ and\ \bibinfo {author} {\bibfnamefont {A.}~\bibnamefont {Vishwanath}},\ }\bibfield  {title} {\bibinfo {title} {Weyl and {D}irac semimetals in three-dimensional solids},\ }\href {https://doi.org/10.1103/RevModPhys.90.015001} {\bibfield  {journal} {\bibinfo  {journal} {Rev. Mod. Phys.}\ }\textbf {\bibinfo {volume} {90}},\ \bibinfo {pages} {015001} (\bibinfo {year} {2018})}\BibitemShut {NoStop}%
\bibitem [{\citenamefont {Fang}\ \emph {et~al.}(2016)\citenamefont {Fang}, \citenamefont {Weng}, \citenamefont {Dai},\ and\ \citenamefont {Fang}}]{fang2016}%
  \BibitemOpen
  \bibfield  {author} {\bibinfo {author} {\bibfnamefont {C.}~\bibnamefont {Fang}}, \bibinfo {author} {\bibfnamefont {H.}~\bibnamefont {Weng}}, \bibinfo {author} {\bibfnamefont {X.}~\bibnamefont {Dai}},\ and\ \bibinfo {author} {\bibfnamefont {Z.}~\bibnamefont {Fang}},\ }\bibfield  {title} {\bibinfo {title} {Topological nodal line semimetals},\ }\href {https://doi.org/10.1088/1674-1056/25/11/117106} {\bibfield  {journal} {\bibinfo  {journal} {Chin. Phys. B}\ }\textbf {\bibinfo {volume} {25}},\ \bibinfo {pages} {117106} (\bibinfo {year} {2016})}\BibitemShut {NoStop}%
\bibitem [{\citenamefont {Luttinger}\ and\ \citenamefont {Kohn}(1955)}]{luttinger1955}%
  \BibitemOpen
  \bibfield  {author} {\bibinfo {author} {\bibfnamefont {J.~M.}\ \bibnamefont {Luttinger}}\ and\ \bibinfo {author} {\bibfnamefont {W.}~\bibnamefont {Kohn}},\ }\bibfield  {title} {\bibinfo {title} {Motion of electrons and holes in perturbed periodic fields},\ }\href {https://doi.org/10.1103/PhysRev.97.869} {\bibfield  {journal} {\bibinfo  {journal} {Phys. Rev.}\ }\textbf {\bibinfo {volume} {97}},\ \bibinfo {pages} {869} (\bibinfo {year} {1955})}\BibitemShut {NoStop}%
\bibitem [{\citenamefont {Bir}\ and\ \citenamefont {Pikus}(1974)}]{bir1974}%
  \BibitemOpen
  \bibfield  {author} {\bibinfo {author} {\bibfnamefont {G.~L.}\ \bibnamefont {Bir}}\ and\ \bibinfo {author} {\bibfnamefont {G.~E.}\ \bibnamefont {Pikus}},\ }\href@noop {} {\emph {\bibinfo {title} {Symmetry and Strain-Induced Effects in Semiconductors}}}\ (\bibinfo  {publisher} {John Wiley \& Sons},\ \bibinfo {address} {New York},\ \bibinfo {year} {1974})\ \bibinfo {note} {translated from the Russian by R. S. Knox}\BibitemShut {NoStop}%
\bibitem [{\citenamefont {Winkler}(2003)}]{winkler2003}%
  \BibitemOpen
  \bibfield  {author} {\bibinfo {author} {\bibfnamefont {R.}~\bibnamefont {Winkler}},\ }\href {https://doi.org/10.1007/b13586} {\emph {\bibinfo {title} {Spin-Orbit Coupling Effects in Two-Dimensional Electron and Hole Systems}}},\ \bibinfo {series} {Springer Tracts in Modern Physics}, Vol.\ \bibinfo {volume} {191}\ (\bibinfo  {publisher} {Springer-Verlag},\ \bibinfo {address} {Berlin},\ \bibinfo {year} {2003})\BibitemShut {NoStop}%
\bibitem [{\citenamefont {Tinkham}(1964)}]{tinkham1964}%
  \BibitemOpen
  \bibfield  {author} {\bibinfo {author} {\bibfnamefont {M.}~\bibnamefont {Tinkham}},\ }\href@noop {} {\emph {\bibinfo {title} {Group Theory and Quantum Mechanics}}},\ International Series in Pure and Applied Physics\ (\bibinfo  {publisher} {McGraw-Hill},\ \bibinfo {address} {New York},\ \bibinfo {year} {1964})\BibitemShut {NoStop}%
\bibitem [{\citenamefont {Hahn}(2005)}]{hahn2005}%
  \BibitemOpen
  \bibinfo {editor} {\bibfnamefont {T.}~\bibnamefont {Hahn}},\ ed.,\ \href {https://doi.org/10.1107/97809553602060000100} {\emph {\bibinfo {title} {International Tables for Crystallography, Volume {A}: Space-Group Symmetry}}},\ \bibinfo {edition} {5th}\ ed.\ (\bibinfo  {publisher} {Kluwer Academic Publishers},\ \bibinfo {address} {Dordrecht},\ \bibinfo {year} {2005})\BibitemShut {NoStop}%
\bibitem [{\citenamefont {Wigner}(1959)}]{wigner1959}%
  \BibitemOpen
  \bibfield  {author} {\bibinfo {author} {\bibfnamefont {E.~P.}\ \bibnamefont {Wigner}},\ }\href@noop {} {\emph {\bibinfo {title} {Group Theory and Its Application to the Quantum Mechanics of Atomic Spectra}}},\ Pure and Applied Physics: A Series of Monographs and Textbooks\ (\bibinfo  {publisher} {Academic Press},\ \bibinfo {address} {New York},\ \bibinfo {year} {1959})\ \bibinfo {note} {translated from the German by J. J. Griffin}\BibitemShut {NoStop}%
\bibitem [{\citenamefont {Sakurai}\ and\ \citenamefont {Napolitano}(2020)}]{sakurai2020}%
  \BibitemOpen
  \bibfield  {author} {\bibinfo {author} {\bibfnamefont {J.~J.}\ \bibnamefont {Sakurai}}\ and\ \bibinfo {author} {\bibfnamefont {J.}~\bibnamefont {Napolitano}},\ }\href {https://www.cambridge.org/9781108473224} {\emph {\bibinfo {title} {Modern Quantum Mechanics}}},\ \bibinfo {edition} {3rd}\ ed.\ (\bibinfo  {publisher} {Cambridge University Press},\ \bibinfo {address} {Cambridge, UK},\ \bibinfo {year} {2020})\BibitemShut {NoStop}%
\bibitem [{\citenamefont {Wieder}\ and\ \citenamefont {Bradlyn}(2021)}]{spacegroupirep}%
  \BibitemOpen
  \bibfield  {author} {\bibinfo {author} {\bibfnamefont {B.~J.}\ \bibnamefont {Wieder}}\ and\ \bibinfo {author} {\bibfnamefont {B.}~\bibnamefont {Bradlyn}},\ }\href {https://github.com/brwieder/spacegroupirep} {\bibinfo {title} {{SpaceGroupIrep}: A {Mathematica} package for irreducible representations of space groups}},\ \bibinfo {howpublished} {GitHub} (\bibinfo {year} {2021}),\ \bibinfo {note} {accessed: 2026}\BibitemShut {NoStop}%
\bibitem [{\citenamefont {Bradlyn}\ \emph {et~al.}(2017)\citenamefont {Bradlyn}, \citenamefont {Elcoro}, \citenamefont {Cano}, \citenamefont {Vergniory}, \citenamefont {Wang}, \citenamefont {Felser}, \citenamefont {Aroyo},\ and\ \citenamefont {Bernevig}}]{bradlyn2017}%
  \BibitemOpen
  \bibfield  {author} {\bibinfo {author} {\bibfnamefont {B.}~\bibnamefont {Bradlyn}}, \bibinfo {author} {\bibfnamefont {L.}~\bibnamefont {Elcoro}}, \bibinfo {author} {\bibfnamefont {J.}~\bibnamefont {Cano}}, \bibinfo {author} {\bibfnamefont {M.~G.}\ \bibnamefont {Vergniory}}, \bibinfo {author} {\bibfnamefont {Z.}~\bibnamefont {Wang}}, \bibinfo {author} {\bibfnamefont {C.}~\bibnamefont {Felser}}, \bibinfo {author} {\bibfnamefont {M.~I.}\ \bibnamefont {Aroyo}},\ and\ \bibinfo {author} {\bibfnamefont {B.~A.}\ \bibnamefont {Bernevig}},\ }\bibfield  {title} {\bibinfo {title} {Topological quantum chemistry},\ }\href {https://doi.org/10.1038/nature23268} {\bibfield  {journal} {\bibinfo  {journal} {Nature}\ }\textbf {\bibinfo {volume} {547}},\ \bibinfo {pages} {298} (\bibinfo {year} {2017})}\BibitemShut {NoStop}%
\bibitem [{\citenamefont {{Bilbao Crystallographic Server}}(1997)}]{bilbao}%
  \BibitemOpen
  \bibfield  {author} {\bibinfo {author} {\bibnamefont {{Bilbao Crystallographic Server}}},\ }\href@noop {} {\bibinfo {title} {{Bilbao Crystallographic Server}}},\ \bibinfo {howpublished} {\url{https://www.cryst.ehu.es/}} (\bibinfo {year} {1997}),\ \bibinfo {note} {accessed: 2026}\BibitemShut {NoStop}%
\bibitem [{\citenamefont {Aroyo}\ \emph {et~al.}(2006)\citenamefont {Aroyo}, \citenamefont {Perez-Mato}, \citenamefont {Capillas}, \citenamefont {Kroumova}, \citenamefont {Ivantchev}, \citenamefont {Madariaga}, \citenamefont {Kirov},\ and\ \citenamefont {Wills}}]{Aroyo2006a}%
  \BibitemOpen
  \bibfield  {author} {\bibinfo {author} {\bibfnamefont {M.~I.}\ \bibnamefont {Aroyo}}, \bibinfo {author} {\bibfnamefont {J.~M.}\ \bibnamefont {Perez-Mato}}, \bibinfo {author} {\bibfnamefont {C.}~\bibnamefont {Capillas}}, \bibinfo {author} {\bibfnamefont {E.}~\bibnamefont {Kroumova}}, \bibinfo {author} {\bibfnamefont {S.}~\bibnamefont {Ivantchev}}, \bibinfo {author} {\bibfnamefont {G.}~\bibnamefont {Madariaga}}, \bibinfo {author} {\bibfnamefont {A.}~\bibnamefont {Kirov}},\ and\ \bibinfo {author} {\bibfnamefont {H.}~\bibnamefont {Wills}},\ }\bibfield  {title} {\bibinfo {title} {{Bilbao Crystallographic Server I}: Databases and crystallographic computing programs},\ }\href {https://doi.org/10.1524/zkri.2006.221.1.15} {\bibfield  {journal} {\bibinfo  {journal} {Z. Kristallogr.}\ }\textbf {\bibinfo {volume} {221}},\ \bibinfo {pages} {15} (\bibinfo {year} {2006})}\BibitemShut {NoStop}%
\bibitem [{\citenamefont {Gao}\ \emph {et~al.}(2021)\citenamefont {Gao}, \citenamefont {Wu}, \citenamefont {Persson},\ and\ \citenamefont {Wang}}]{GAO2021107760}%
  \BibitemOpen
  \bibfield  {author} {\bibinfo {author} {\bibfnamefont {J.}~\bibnamefont {Gao}}, \bibinfo {author} {\bibfnamefont {Q.}~\bibnamefont {Wu}}, \bibinfo {author} {\bibfnamefont {C.}~\bibnamefont {Persson}},\ and\ \bibinfo {author} {\bibfnamefont {Z.}~\bibnamefont {Wang}},\ }\bibfield  {title} {\bibinfo {title} {{IRVSP}: To obtain irreducible representations of electronic states in the {VASP}},\ }\href {https://doi.org/10.1016/j.cpc.2020.107760} {\bibfield  {journal} {\bibinfo  {journal} {Comput. Phys. Commun.}\ }\textbf {\bibinfo {volume} {261}},\ \bibinfo {pages} {107760} (\bibinfo {year} {2021})},\ \Eprint {https://arxiv.org/abs/2011.06945} {arXiv:2011.06945 [cond-mat.mtrl-sci]} \BibitemShut {NoStop}%
\bibitem [{\citenamefont {Herath}\ \emph {et~al.}(2020)\citenamefont {Herath}, \citenamefont {Tavadze}, \citenamefont {He}, \citenamefont {Bousquet}, \citenamefont {Singh}, \citenamefont {Muñoz},\ and\ \citenamefont {Romero}}]{HERATH2020107080}%
  \BibitemOpen
  \bibfield  {author} {\bibinfo {author} {\bibfnamefont {U.}~\bibnamefont {Herath}}, \bibinfo {author} {\bibfnamefont {P.}~\bibnamefont {Tavadze}}, \bibinfo {author} {\bibfnamefont {X.}~\bibnamefont {He}}, \bibinfo {author} {\bibfnamefont {E.}~\bibnamefont {Bousquet}}, \bibinfo {author} {\bibfnamefont {S.}~\bibnamefont {Singh}}, \bibinfo {author} {\bibfnamefont {F.}~\bibnamefont {Muñoz}},\ and\ \bibinfo {author} {\bibfnamefont {A.~H.}\ \bibnamefont {Romero}},\ }\bibfield  {title} {\bibinfo {title} {{PyProcar}: A python library for electronic structure pre/post-processing},\ }\href {https://doi.org/10.1016/j.cpc.2019.107080} {\bibfield  {journal} {\bibinfo  {journal} {Computer Physics Communications}\ }\textbf {\bibinfo {volume} {251}},\ \bibinfo {pages} {107080} (\bibinfo {year} {2020})}\BibitemShut {NoStop}%
\bibitem [{\citenamefont {Lang}\ \emph {et~al.}(2024)\citenamefont {Lang}, \citenamefont {Tavadze}, \citenamefont {Tellez}, \citenamefont {Bousquet}, \citenamefont {Xu}, \citenamefont {Muñoz}, \citenamefont {Vasquez}, \citenamefont {Herath},\ and\ \citenamefont {Romero}}]{LANG2024109063}%
  \BibitemOpen
  \bibfield  {author} {\bibinfo {author} {\bibfnamefont {L.}~\bibnamefont {Lang}}, \bibinfo {author} {\bibfnamefont {P.}~\bibnamefont {Tavadze}}, \bibinfo {author} {\bibfnamefont {A.}~\bibnamefont {Tellez}}, \bibinfo {author} {\bibfnamefont {E.}~\bibnamefont {Bousquet}}, \bibinfo {author} {\bibfnamefont {H.}~\bibnamefont {Xu}}, \bibinfo {author} {\bibfnamefont {F.}~\bibnamefont {Muñoz}}, \bibinfo {author} {\bibfnamefont {N.}~\bibnamefont {Vasquez}}, \bibinfo {author} {\bibfnamefont {U.}~\bibnamefont {Herath}},\ and\ \bibinfo {author} {\bibfnamefont {A.~H.}\ \bibnamefont {Romero}},\ }\bibfield  {title} {\bibinfo {title} {Expanding {PyProcar} for new features, maintainability, and reliability},\ }\href {https://doi.org/10.1016/j.cpc.2023.109063} {\bibfield  {journal} {\bibinfo  {journal} {Computer Physics Communications}\ }\textbf {\bibinfo {volume} {297}},\ \bibinfo {pages} {109063} (\bibinfo {year} {2024})}\BibitemShut {NoStop}%
\bibitem [{\citenamefont {Wieder}\ \emph {et~al.}(2021)\citenamefont {Wieder}, \citenamefont {Bradlyn}, \citenamefont {Schoop}, \citenamefont {Topp},\ and\ \citenamefont {Cava}}]{tomsg}%
  \BibitemOpen
  \bibfield  {author} {\bibinfo {author} {\bibfnamefont {B.~J.}\ \bibnamefont {Wieder}}, \bibinfo {author} {\bibfnamefont {B.}~\bibnamefont {Bradlyn}}, \bibinfo {author} {\bibfnamefont {L.~M.}\ \bibnamefont {Schoop}}, \bibinfo {author} {\bibfnamefont {A.}~\bibnamefont {Topp}},\ and\ \bibinfo {author} {\bibfnamefont {R.~J.}\ \bibnamefont {Cava}},\ }\href {https://github.com/brwieder/ToMSGKpoint} {\bibinfo {title} {{ToMSGKpoint}: Representation analysis for magnetic space groups}},\ \bibinfo {howpublished} {GitHub} (\bibinfo {year} {2021}),\ \bibinfo {note} {accessed: 2026}\BibitemShut {NoStop}%
\bibitem [{\citenamefont {Dresselhaus}\ \emph {et~al.}(2008)\citenamefont {Dresselhaus}, \citenamefont {Dresselhaus},\ and\ \citenamefont {Jorio}}]{dresselhaus2008}%
  \BibitemOpen
  \bibfield  {author} {\bibinfo {author} {\bibfnamefont {M.~S.}\ \bibnamefont {Dresselhaus}}, \bibinfo {author} {\bibfnamefont {G.}~\bibnamefont {Dresselhaus}},\ and\ \bibinfo {author} {\bibfnamefont {A.}~\bibnamefont {Jorio}},\ }\href {https://doi.org/10.1007/978-3-540-32899-5} {\emph {\bibinfo {title} {Group Theory: Application to the Physics of Condensed Matter}}},\ \bibinfo {series} {Springer Series in Solid-State Sciences}, Vol.\ \bibinfo {volume} {175}\ (\bibinfo  {publisher} {Springer-Verlag},\ \bibinfo {address} {Berlin},\ \bibinfo {year} {2008})\BibitemShut {NoStop}%
\bibitem [{\citenamefont {Shtyk}\ \emph {et~al.}(2017)\citenamefont {Shtyk}, \citenamefont {Goldstein},\ and\ \citenamefont {Chamon}}]{shtyk2017}%
  \BibitemOpen
  \bibfield  {author} {\bibinfo {author} {\bibfnamefont {A.}~\bibnamefont {Shtyk}}, \bibinfo {author} {\bibfnamefont {G.}~\bibnamefont {Goldstein}},\ and\ \bibinfo {author} {\bibfnamefont {C.}~\bibnamefont {Chamon}},\ }\bibfield  {title} {\bibinfo {title} {Electrons at the monkey saddle: A multicritical lifshitz point},\ }\href {https://doi.org/10.1103/PhysRevB.95.035137} {\bibfield  {journal} {\bibinfo  {journal} {Phys. Rev. B}\ }\textbf {\bibinfo {volume} {95}},\ \bibinfo {pages} {035137} (\bibinfo {year} {2017})}\BibitemShut {NoStop}%
\bibitem [{\citenamefont {Zhang}\ \emph {et~al.}(2022)\citenamefont {Zhang}, \citenamefont {Yu}, \citenamefont {Liu},\ and\ \citenamefont {Yao}}]{zhang2022}%
  \BibitemOpen
  \bibfield  {author} {\bibinfo {author} {\bibfnamefont {Z.}~\bibnamefont {Zhang}}, \bibinfo {author} {\bibfnamefont {Z.-M.}\ \bibnamefont {Yu}}, \bibinfo {author} {\bibfnamefont {G.-B.}\ \bibnamefont {Liu}},\ and\ \bibinfo {author} {\bibfnamefont {Y.}~\bibnamefont {Yao}},\ }\bibfield  {title} {\bibinfo {title} {{MagneticTB}: A package for tight-binding model of magnetic and non-magnetic materials},\ }\href {https://doi.org/10.1016/j.cpc.2021.108153} {\bibfield  {journal} {\bibinfo  {journal} {Comput. Phys. Commun.}\ }\textbf {\bibinfo {volume} {270}},\ \bibinfo {pages} {108153} (\bibinfo {year} {2022})}\BibitemShut {NoStop}%
\end{thebibliography}%

\end{document}